\documentclass[11pt]{article}
\usepackage{amsmath,latexsym,amssymb,cite}

\def\baselinestretch{1.4}

\makeatletter
\@addtoreset{equation}{section}
\makeatother

\def\ben{\begin{equation}}
\def\een{\end{equation}}
\def\ee{\een}
\def\be {\ben}
\def\half{{1 \over 2}}
\def\bea{\begin{eqnarray}}
\def\eea{\end{eqnarray}}
\def\bx {{\bf x}}
\def\br {{\bf r}}
\def\bk {{\bf k}}
\def \p {\partial}
\def \nn {\nonumber}
\def \fft  {\frac}
\def\Box{\square}

\def \ft12{\half}

\def\be{\begin{eqnarray}}
\def\ee{\end{eqnarray}}

\def\>{\rangle}
\def\<{\langle}

\def\lb{\label}

\def\nn{\nonumber}
\def\p{\partial}

\def\del{\partial}
\def\cL{{{\cal L}}}

\def\0{{\sst{(0)}}}
\def\1{{\sst{(1)}}}
\def\2{{\sst{(2)}}}
\def\3{{\sst{(3)}}}
\def\4{{\sst{(4)}}}
\def\5{{\sst{(5)}}}
\def\6{{\sst{(6)}}}
\def\7{{\sst{(7)}}}
\def\8{{\sst{(8)}}}
\def\sst#1{{\scriptscriptstyle #1}}

\newcount\hour \newcount\minute
\hour=\time  \divide \hour by 60
\minute=\time
\loop \ifnum \minute > 59 \advance \minute by -60 \repeat
\def\nowtwelve{\ifnum \hour<13 \number\hour:
                      \ifnum \minute<10 0\fi
                      \number\minute
                      \ifnum \hour<12 \ A.M.\else \ P.M.\fi
         \else \advance \hour by -12 \number\hour:
                      \ifnum \minute<10 0\fi
                      \number\minute \ P.M.\fi}
\def\nowtwentyfour{\ifnum \hour<10 0\fi
                \number\hour:
                \ifnum \minute<10 0\fi
                \number\minute}

\newcommand{\hoch}[1]{$\, ^{#1}$}

\newcommand{\auth}{\large\bf{G.W. Gibbons\hoch{1}, C.N. Pope\hoch{2,1}
and Sergey Solodukhin\hoch{3}}}

\begin{document}
\begin{flushright}
\hfill {MI-TH-1927}\\
\end{flushright}

\begin{center}

{\bf\large Higher Derivative Scalar Quantum Field Theory in Curved Spacetime}

\vspace{15pt}

\auth

{\small

\vspace{8pt}{\hoch{1}\it DAMTP, Centre for Mathematical Sciences,\\
 Cambridge University, Wilberforce Road, Cambridge CB3 OWA, UK}

\vspace{4pt}{\hoch{2}\it Mitchell 
Institute for Fundamental
Physics and Astronomy,\\
Texas A\&M University, College Station, TX 77843-4242, USA
}

\vspace{4pt}{\hoch{3}\it Institut Denis Poisson, UMR CNRS 7013,,
Universit\'e de Tours, Universit\'e d'Orl\' eans,\\
Parc de Grandmont, 37200 Tours, France}

}

\vspace{20pt}

\underline{\small ABSTRACT}
\end{center}

\vspace{5pt}

We study a free scalar field $\phi$ in a fixed curved background spacetime
subject to a higher derivative field equation of the
form $F(\Box)\phi =0$, where $F$ is a polynomial of the form 
$F(\Box)= \prod_i (\Box-m_i^2)$ and all masses $m_i$ are distinct and real.
Using an auxiliary field method to simplify the calculations, 
we obtain expressions
for the Belinfante-Rosenfeld symmetric energy-momentum tensor
and compare it with the canonical energy-momentum tensor when the background is 
Minkowski spacetime. We also obtain the conserved symplectic current
necessary for quantisation and briefly discuss the issue of
negative energy versus negative norm and its relation to Reflection Positivity
in Euclidean treatments. We study, without assuming spherical symmetry,
the possible existence of finite energy static solutions of the
scalar equations, in static or stationary background geometries. 
Subject to various assumptions on the potential, we establish 
non-existence results including a no-scalar-hair 
theorem for static black holes.  We consider Pais-Uhlenbeck field theories
in a cosmological de Sitter background, and show how the Hubble friction
may eliminate what would otherwise be unstable behaviour when interactions
are included.

\medskip

\pagebreak

\def\baselinestretch{1.3}
\tableofcontents
\def\baselinestretch{1.4}

\section{Introduction}

  Higher-derivative field theories have received a considerable amount of
attention over the years for a variety of reasons, not least because of the
realisation that theories incorporating standard Einstein gravity inevitably
suffer from problems of non-renormalisability.  Although it was
shown by 't Hooft and Veltman \cite{'tHooft:1974bx} that pure
Einstein gravity itself was finite at one loop, this does not persist
at higher loop order \cite{Goroff:1985sz}, nor if (generic) matter is included.  
The demonstration by Stelle \cite{Stelle:1976gc} in the 1970s that the 
non-renormalisability problem itself could be
overcome by adding quadratic curvature terms to the action opened the
door to many investigations of such theories, although the
realisation that renormalisability could only thereby be achieved
at the price of introducing ghost states of negative norm, or energies
that are unbounded below, dampened the
enthusiasm for the idea.  Subsequently, with the development of string theory, 
the idea of higher-derivative corrections to gravity acquired a new impetus,
but now in the framework of an effective field theory in which quadratic
curvature corrections would represent just the start of an infinite 
sequence of higher-order terms.  Within this framework, a focus on any
particular lower-order term or terms in the infinite sequence of corrections
would seem to be unjustified, since in circumstances where such a term or
terms leads to substantial modifications to the solutions or spectrum of the
theory, yet-higher order terms that are being ignored would have at least
as important an effect.  

Although the conviction that the ghost states or unbounded negative energies
of a finite-order higher 
derivative theory are sufficient to rule them out of consideration
is widespread, it is perhaps still worthwhile to study them in more detail,
and to investigate whether the ostensible problems really are as severe
as is commonly believed. 
One of the earliest investigations of higher-derivative field theories was
in the classic work of Pais and Uhlenbeck \cite{Pais:1950za}, who studied
scalar field theories in Minkowski spacetime.  Much of their focus was
on differential operators of infinite order, such as the exponential of
the d'Alembertian, but they did also consider operators of finite order,
such as powers of the d'Alembertian or products of massive Klein-Gordon
operators.  Many of the features that are encountered in theories such as these
will have counterparts in more complicated theories such as higher-derivative
theories of gravity.  

  As a preliminary to studying higher-derivative field theories, much insight
can be gained in the simpler case of a classical higher-derivative particle
mechanical
theory, and its quantum mechanical extension.  A simple and much studied
example is the Pais-Uhlenbeck oscillator, $(D^2-\omega_1^2)(D^2-\omega_2^2)x(t)=0$,
where $D=d/dt$.  Its Hamiltonian, constructed using the Ostrogradsky
procedure \cite{Ostrogradsky:1850fid}, which corresponds essentially to the
difference between standard oscillator Hamiltonians for particles with
frequencies $\omega_1$ and $\omega_2$, is unbounded both above and below.
At the level of the free theory this is not a problem, but a widespread intuition is
that once interactions are included, instabilities will inevitably set in since
the energies in the 
negative energy modes and the positive energy modes can both increase (in magnitude)
while the total energy remains constant.\footnote{Although this intuition is
commonly attributed to Ostrogradsky, either as the ``Ostrogradsky Instability''
or ``Ostrogradsky's Instability Theorem,'' it appears that Ostrogradsky himself
never actually 
discussed the question of instabilities in higher-derivative theories.} 
A number of studies have show that
this need not in fact be the case \cite{smilga1,smilga2,Smilga:2008pr,pavsic1}.  
For example, for given
initial conditions, an added $\lambda x^4$ interaction term will not give 
instabilities provided $\lambda$ is sufficiently small \cite{pavsic1}.  And if
one instead adds a bounded potential, such as $\lambda \sin^4 x$, the evolution
of initial data is unconditionally stable \cite{pavsic1}.  For a 
consideration of of the Pais-Uhlenbeck oscillator from the point of view
of statistical physics, see \cite{nicolis}.

  It has often been argued (see, for example, \cite{Woodard}) that instability 
problems in a higher-derivative field theory will be much more severe than
in a particle mechanical theory, because of the diverging entropy associated with the
infinity of unobserved high-momentum states, and that this would lead to 
an instantaneous
decay of the vacuum.  Countering this, it has been argued that the divergent
entropy results from integrating over the phase space of unobserved decay
products, and that in a closed system such as our universe there exist no
external observers who would integrate over this infinite phase space. 
Actual observers, inside the universe, would implicitly have measured the
momenta of the particles, and so there would be no diverging phase space integrals
leading to instantaneous vacuum decay \cite{pavsic2}.   It has, further, been
argued that the experience with adding interactions to a higher-derivative
particle mechanics model, where potentials that are bounded both above and below
do not give rise to any instabilities, suggests that similarly bounded potentials in
a higher-derivative field theory may not in any case give rise to instabilities
\cite{pavsic2}.  There have also been proposals,
such as that by Hawking \cite{Hawking:1985gh}, that one might impose 
boundary conditions to eliminate ghost excitations, at least in asymptotic
states, since the concomitant acausality could be at sufficiently 
short time scales that it may not be observable in a Wheeler-DeWitt framework. 
More recently, Hawking and Hertog \cite{Hawking:2001yt} 
argued that in a path-integral
treatment in a Euclidean framework, one can obtain well-defined rules for
calculating probabilities, after continuing to Lorentzian spacetime, even in a
higher-derivative field theory.

   In the light of some of these investigations, we shall take the view that
it is still of considerable interest to study the properties of 
higher-derivative
field theories, and that scalar theories of this kind may be useful 
as models for
more complicated systems such as higher-derivative gravities.  Most of the 
previous work on higher-derivative scalar field theories has been in a flat
Minkowski spacetime background. 
 Our main purpose, in this paper, is to consider scalar theories of the kind 
studied by
Pais and Uhlenbeck, but now within the framework of curved spacetime 
backgrounds.
We shall show how the calculation of the Belinfante-Rosenfeld 
energy-momentum tensor can be greatly
simplified by introducing an auxiliary field formalism, and we shall also make
comparisons with the canonical flat spacetime energy-momentum tensor 
calculated using Noether methods.  We also obtain some general results
for the symplectic Noether currents that can be used in order to 
construct the norms on states in the theories.  One further motivation for
studying higher-derivative scalar field theories in curved backgrounds is
that, as we shall discuss later, this can actually help to mitigate some
of the instabilities that might otherwise occur in flat spacetime.

  The organisation of the paper is as follows. We begin in section 2 with
a review of some basic aspects of the classical and quantum properties of
a point particle moving in one dimension, governed by a higher-derivative 
equation of motion.  Our discussion includes a review of the Ostrogradsky 
\cite{Ostrogradsky:1850fid,Whittaker} construction of the Hamiltonian 
describing 
the system, and an elegant analysis by Smilga \cite{Smilga:2008pr} of the 
quantum theory of a 4th-order Pais-Uhlenbeck oscillator.  In section 3 
we review some properties of a scalar Pais-Uhlenbeck field theory in 
a Minkowski background, including the construction of the canonical 
energy-momentum tensor and the conserved symplectic current.
In section 4 we extend our discussion to the scalar Pais-Uhlenbeck field
theory in a curved spacetime background, showing how one can use an 
auxiliary field formulation in order to facilitate the construction of the
Belinfante-Rosenfeld energy-momentum tensor, which allows a 
consistent coupling to the gravitational 
field. By this means, one avoids the necessity of varying the metrics in
high powers of the covariant derivative.

In section 5 we discuss the quantisation of higher
derivative scalar field theories using Euclidean  methods.
This may be done if the background metric admits an
analytic continuation containing a real section on which the metric
is positive definite, as it does for real tunnelling metrics
\cite{Gibbons:1990ns}. It may  also arise in a full blown
Euclidean quantum gravity path integral
calculation when a  gravitational instanton saddle point
admits a reflection map,
i.e. an involutive isometry fixing a separating hypersurface.
We generalise previous arguments \cite{Hawking:1985gh}  showing that
if one adopts
the Osterwalder and Schrader prescription \cite{Osterwalder:1973dx} 
for constructing the quantum mechanical Hilbert space, then
Reflection Positivity is not satisfied
despite the Euclidean action of the scalar field
being positive definite. In other words,
negative norm states are inevitable.

In section 6, subject to various assumptions about the
scalar potential, if present, we prove non-existence results for static
solutions of the equations of motion in static or stationary
spacetimes.
The background could be Minkowski spacetime, in which case
these results rule out stable solitons and unstable sphaleron type solutions.
They also hold in globally stationary backgrounds. They are
easily extended to cover
the case of static black hole backgrounds with regular event horizons.

In section 7, we consider Pais-Uhlenbeck field theories in a de Sitter
spacetime background, and show how Hubble friction can eliminate instabilities
that would otherwise occur when nonlinear couplings are present.  These
examples serve to illustrate the fact that the often-claimed instabilities
of higher-order field theories can sometimes be illusory.

\section{A Higher Derivative Point Particle}

To orient ourselves, we begin by recalling some facts about
theories of particles moving in one spatial dimension governed
by a Lagrangian $L=L(x,\dot x ,\ddot x,t) $  depending upon the 
position $x$, velocity $\dot x$ and acceleration $\ddot x$. 
Varying the Lagrangian gives
\bea
\delta L -    \delta t \frac{\p L}{\p t}  &=&
\delta x \frac{\p L}{\p x} + \delta \dot x \frac{\p L}{\p \dot x}  
+ \delta \ddot x \frac{\p L}{\p \ddot x} \\  
&=& \delta x ( \frac{\p L}{\p x} - \frac{d}{dt} \frac{\p L}{\p \dot x} )
+ \frac{d }{dt} ( \delta x \frac{\p L}{\p \dot x } ) 
 - \delta \dot x \frac{d }{dt}  \frac {\p L}{\p \ddot x}  + \frac{d}{dt} 
 ( \delta \dot x \frac{\p L}{\p \ddot x}  ) \\  
&=&  \delta x (\frac{\p L}{\p x} - \frac{d}{dt} \frac{\p L}{\p \dot x} + 
\frac{d ^2}{dt ^2} \frac{\p L}{\p \ddot x}  )
+ \frac{d}{dt}( \delta x \frac{\p L}{\p \dot x}  + \delta \dot x 
\frac{\p L}{\p \ddot x}   - \delta x \frac{d}{dt}  \frac{\p L}{\p \ddot x})\,.    
\eea       
The equation of motion is thus
\ben
\frac{\p L}{\p x} - \frac{d}{dt} \frac{\p L}{\p \dot x} + 
\frac{d ^2}{dt ^2} \frac{\p L}{\p \ddot x} =0 \,.
\label{EOM0} 
\een

\subsection{Energy and and Momentum Conservation}

If the equation of motion holds, then
\ben
\frac {d}{dt} \Big(  L - \dot  x \frac{\p L}{\p \dot x}  -  \ddot x 
\frac{\p L}{\p \ddot x}   + \dot x  \frac{d}{dt}  \frac{\p L}{\p \ddot x}
\Big) = \frac{\p L}{\p t}\,.  
\een
Thus, if $L$ does not depend upon $t$, there is a conserved energy 
\ben
E=   \dot  x \, \frac{\p L}{\p \dot x}  +  \ddot x\, 
\frac{\p L}{\p \ddot x}   - \dot x \, \frac{d}{dt}  
\frac{\p L}{\p \ddot x}-L\,,
\een
which may be written as 
\ben
E=   \dot x \,\Big( \frac{\p L}{\p \dot x}  - 
\frac{d}{dt}  \frac{\p L}{\p \ddot x}\Big)       +  \ddot x\, 
\frac{\p L}{\p \ddot x}    -L \,.
\een

If the Lagrangian is translationally  invariant, $\frac{\p L}{\p x}=0$, 
we expect that momentum should be conserved.
With an eye on the equation of motion (\ref{EOM0}), we define the 
momentum $p_x$  by 
\ben
p_x= \frac{\p L}{\p \dot x} -\frac{d}{dt} \frac{\p L}{\p \ddot x } \,,
\een  
so that  equation of motion (\ref{EOM0}) takes the form
\ben 
\dot p_x = \frac{\p L}{\p x}\,,
\een 
and hence if the Lagrangian is translationally invariant
the momentum  $p_x$ is conserved.  In terms of $p_x$, the energy is given by 
\ben
E= \dot x \, p_x + \ddot x \, \frac{\p L} {\p \ddot x} - L \,.   
\een

\subsection{Ostrogradsky's Hamiltonian}  

 Following
\cite{Ostrogradsky:1850fid,Whittaker,Pais:1950za}\footnote{For a geometrical
treatment, see \cite{mavicica}.},
we define
\ben
y= \dot x \,,\qquad  p_y  = \frac{\p L}{\p \ddot x} \,.
\een
If the ``nondegeneracy'' condition
\ben
\frac{\p ^2 L}{\p \ddot x ^2 } \ne 0
\een
holds, then we may solve for the acceleration as a function of $(x,y,p_y)$:  
\ben
\ddot x = a(x,y,p_y)\,.  
\een
To obtain the Hamiltonian, we take the Legendre transform of
the Lagrangian and obtain    
\ben
H= p_x \,y + p_y \, a(x,y, p_y) - L(x,y, a(x,y,p_y )) \,.
\een

  As a concrete example, consider the Pais-Uhlenbeck oscillator, described by the
Lagrangian
\ben
L = \half \dot x^2 - \half \beta \,\ddot x^2 - \half \Omega^2\, x^2\,.  
\een
Clearly the  non-degeneracy condition holds holds as long as $\beta \ne 0$, 
since   
\ben
\frac{\p ^2 L}{\p \ddot x^2 }= -\beta \,.
\een

The equations of motion are 
\ben
\ddot x + \beta \, \ddddot x   + \Omega^2\, x =0 \,,  
\een
i.e., 
\ben
p_y = -\beta\, \ddot x \,, \qquad \Rightarrow \qquad \ddot x= -
\frac{1}{\beta} \, p_y = a \,.    
\een
We have 
\ben
p_x= \dot x + \beta \,\, \dddot x \,,
\een
and so the equation of motion implies 
\ben
\dot p_x =- \Omega^2 \,x \,.
\een
In a general potential $V(x)$ we would have 
\ben
\dot p_x= -\frac{dV}{dx} \,.
\een 
Thus \emph{Newton's second law, expressed in terms of momentum is obeyed}.

Making the ansatz $x= \Re  A e^{-i\omega t}$ we find 
\ben
\omega ^4 -  \frac{1}{\beta} \, \omega ^2 + \frac{\Omega ^2}{\beta} =0\,,
\een 
and thus
\ben
\omega^2 =   \frac{1}{2\beta }\bigl( 1  \pm \sqrt{ 1 -4 \beta\, \Omega ^2 } 
\bigr )\,. 
\een 
If $0<4\beta\, \Omega^2<1$, 
then all four roots 
$\omega = (\pm \omega _1, \pm \omega _2)$   will be real, distinct and 
non-vanishing . 
Useful relations are 
\ben
\omega _1^2 + \omega _2 ^2 =  \frac{1}{\beta} \,,
\qquad \omega _1^2 \omega_2^2
=  \frac {\Omega ^2 } {\beta} \,.      
\een

The equation of motion may be written in terms of $\omega_1$ and $\omega_2$ as 
\ben
\ddddot  x   + (\omega_1^2  + \omega _2^2  )  \ddot x  + 
\omega _1^2 \omega_2 ^2 x=0\,.  
\een

We write the general solution as 
\ben
x= A e^{-i\omega_1 t} + B e^ {-i\omega _2 t} + 
\bar A   e^{i\omega_1 t} +\bar B e^ {i\omega _2 t}\,,
\een 
and therefore
\bea
x(0) &=& (A + \bar A) +  (B   + \bar B) \,,\nn \\
\dot x(0) &=& -i \omega _1  (A -\bar A) - i\omega _2 (B-\bar B) \,,\nn\\   
\ddot x(0) &=& -\omega _1 ^2 (A + \bar A) - \omega_2 ^2 (B+ \bar B) \,,\nn\\ 
\dddot x(0) &=&  i \omega _1^3 ( A-\bar A) +i \omega _2 ^3 (B-\bar B)  \,.
\eea
Since $\omega _1\ne \omega _2$, there  exits  a unique solution
for $A$ and $B$ for all initial data $(x(0),\dot x(0), \ddot x(0)$ and 
$\dddot x (0) )$.
Moreover, this solution is bounded for all time. 

\emph {Thus the theory is predictive, the initial value problem has a solution
for all time and it is well posed: it  depends continuously 
on the initial data.}

It is, however, true that the conserved energy    
\ben
E= \half (\dot x) ^2 - \half \beta\, (\ddot x)^2 + 
 \beta \,\dot x \,\dddot x  + \half \Omega^2\, x^2 \,,   
\een
or equivalently in terms of the canonically conjugate variables,
the Hamiltonian 
\bea
H&=& p_x \, y - \frac{p_y ^2}{2 \beta} - \half y^2 + \half \Omega^2 \,x^2\nn\\
&=&  \half p_x ^2 +\half \Omega ^2 x^2 - \half (y- p_x )^2 +
\frac{p_y^2}{2 \beta} \,,    
\eea
depends linearly on $p_x$, and it is the sum of two positive and two negative
squares (if $\beta$ is positive, as discussed above). 
In fact Hamilton's equations  
\bea
\dot p_x =- \Omega ^2 x \,,&\qquad& \dot p_y = -p_x +y \,,\nn\\
\dot  x= y \,, &\qquad & \dot y = -\frac{p_y}{\beta} \label{Heq}
\eea
in this case are linear, and one may verify that they have the same 
characteristic frequencies.    

As mentioned in the introduction, 
it is less obvious what happens if an interaction term
is added, rendering the equations of motion non-linear.
An example would be a potential term of the form $\lambda x^4$, where $\lambda$
is a constant.   
A widely held intuition is that some sort of instability will result,
but, as shown in \cite{pavsic1}, for given initial conditions
the evolution may remain bounded if the $\lambda$ 
interaction is sufficiently small.  The question may also 
depend crucially on the definition
of stability.  The motion may always remain within a 
bounded neighbourhood of
the unperturbed motion, $x=0$ for example, even though it 
may not return asymptotically to $x=0$. 
Moreover the time scale for any
instability may, in the cosmological context, be sufficiently long,
 for small enough $\lambda$, 
as not to lead to any observable effect within the relevant time scale.

\subsection{Quantisation of the Pais-Uhlenbeck Oscillator}

An obvious procedure would be  
to consider wave functions of the form $\Psi=\Psi (x,y)$, in which case 
\ben
\hat p_x= -i \frac{\p}{\p x} \,,\qquad \hat p_y  = -i \frac{\p}{\p y} \,,  
\een
 leading to
\ben
\hat H = -i y \frac{\p}{\p x} - \ft12 \frac{ \p^2}{\p y^2} -
 \half y^2    +        \half \Omega^2\, x ^2 
\,.
\een
More revealing would be  to consider wave functions of the form $\Psi=\Psi 
(x,p_y) $ in which case 
\ben
\hat p=-i \frac{\p}{\p x}\,,\qquad \hat y = i \frac{\p}{\p p_y}\,,
\een
leading to
\ben
\hat H = \frac{\p ^2}{\p x \p y} - \half \frac{\p ^2}{\p p_y ^2 } 
- \frac{p_y  ^2 }{2 \beta} + \half \Omega^2\,  x^2 \,. 
\een
The first two terms  are the wave equation in two-dimensional
Minkowski spacetime with the flat metric
\ben
ds ^2 = dx^2 + 2 dx \, d p_y  = (dx+ d p_y )^2  - d p_y ^2 \,,
\een  
in which $x$ is a lightlike coordinate. 

   Smilga \cite{Smilga:2008pr} succeeded in finding a canonical transformation
which clarifies what is going on.  Starting with the (rescaled) Lagrangian 
\ben
L= \half  \Bigl[ (\ddot x)^2  - (\omega _1^2 + \omega _2^2) (\dot x) ^2
+\omega _1^2  \omega _2 ^2  x^2 \Bigr] \,.
\een
Ostrogradsky's Hamiltonian becomes  
\ben
H=p_x\, y + \half p_y ^2 +  \half (\omega _1^2 + \omega _2^2 )\,y^2  - 
\half \omega_1^2 \, \omega_2^2 \, x^2 \,.   
\een

Now let 
\bea
x &=& \frac{1}{\omega_1} \frac{\omega _1 \,Y- p_X }{\sqrt{\omega_1^2 - 
  \omega_2^2 }} \,,\qquad
p_x= \omega_1 \frac{\omega_1\, p_Y- \omega_2^2 \,X }
{\sqrt{\omega _1^2 - \omega _2^2 }}\,,\nn\\
y &=& 
\frac{\omega_1 \,X- p_Y }{\sqrt{\omega_1^2 - \omega_2^2 }} \,,\qquad \,
p_y=\frac{\omega_1\, p_X- \omega_2^2 \,Y}{\sqrt{\omega_1^2 - \omega_2^2 }} \,,
\eea
which implies
\bea 
X&=&\frac{1}{\omega_1} \frac {p_x+ \omega_1^2\, y }
{\sqrt{\omega_1^2 -\omega_2^2}}\,,\qquad Y= 
\frac{p_y + \omega_1^2\,  x}{\sqrt{\omega_1^2 - \omega_2^2}}\,,\nn\\  
p_X &=& \omega_1 \frac{p_y+\omega_2^2\, x}{\sqrt{\omega_1^2 - \omega_2^2 }}
 \,,\qquad p_Y = \frac{p_x+\omega^2_2\, y}{\sqrt{\omega_1^2-\omega_2^2 }}\,.
\eea
This is a canonical transformation, since
\ben
dp_x \wedge dx + d p_y \wedge dy = dp_X \wedge dX + dp_Y \wedge dY \,.  
\een
Pulling back the Hamiltonian one finds
\ben
H= \half\Bigl(p_X^2 +  \omega_1^2\, X^2 \Bigr ) -  
\half \Bigl(p_Y^2 +  \omega_2^2\, Y^2 \Bigr ) \,. 
\een
This may be quantised in the obvious way, with a positive norm on
the Hilbert space, by introducing annihilation and creation operators 
in the standard way, with the non-vanishing commutators
$[a_X,a_X^\dagger]=[a_Y,a_Y^\dagger]=1$. We have 
\ben
\hat H = ({{\hat a}_X}^\dagger\, {\hat a}_X  +\half   ) \,\omega_1 
  - ({{\hat a}_Y}^\dagger\, {\hat a}_Y  +\half   )\, \omega_2\,.
\een
Assuming that the ``ground state'' satisfies
\ben
{\hat a}_X \,| 0\rangle=  {\hat a}_Y \,| 0\rangle =0 \,, 
\een
all norms are positive, but the spectrum of the Hamitonian is 
unbounded below. 

Alternatively, we could  assume that the vacuum is defined by
\ben
{\hat a}_X \,| 0\rangle=  { {\hat a}_Y}^\dagger\, | 0\rangle =0 \,, 
\een
in which case  the spectrum of the Hamitonian is bounded below
but the norm on states would be indefinite.  For example, the state
$a_Y\, |0\rangle$ now has norm
\be
|a_Y\,|0\rangle|^2= \langle 0|a_Y^\dagger\, a_Y\, |0\rangle =
  -\langle 0|0\rangle\,.
\ee

\section{Pais-Uhlenbeck Field Theory in Minkowski Spacetime} 

In this section we recall some aspects of
the simplest non-interacting higher derivative
scalar field theories in a Minkowski background spacetime.
The  field equations are of the form \cite{Pais:1950za}  
\ben
E\phi \equiv F(\Box)\phi=0 \,,  \label{Pais} 
\een
where\footnote{Throughout the paper we 
adhere to the same $-+++$ signature convention as  \cite{Pais:1950za}.} 
\ben
\Box= -\p^2_t+\nabla ^2 \, = \eta ^{\mu \nu}\p_\mu \p_\nu \,, \qquad  
\p_\mu = (\p_t,\p_{x^i} ) \,
\een
and $F(u)$ is a real valued function of $u$.

A suitable action for a real scalar $\phi$, up to factors and 
possible boundary terms, is  
\ben
\fft12 \int d^4 x\,  \phi F(\Box) \phi\,. \label{action} 
\een
The   Lagrangian models  studied by Ostrogradsky 
are  obtained by replacing $\Box \phi $  in (\ref{Pais}) 
by $-D^2$, where   $D= \frac{d }{dt} $.

   A simple non-trivial example is
\ben
E \phi =(\Box - m_1^2) (\Box - m_2^2 )\phi =0 \label{non}\,,
\een
where, unless otherwise stated, we assume that the masses $m_1$ and $m_2$ are 
real and distinct. The case when $m_1^2=m_2^2$ has some different properties,
and  is referred to as  di-polar.    
If $\Box=-D^2$ we obtain the so-called  {\it Pais-Uhlenbeck oscillator}.

 As we shall see, one may always introduce a symplectic current 
$J^\mu (\phi_1,\phi_2)= - J^\mu(\phi_2, \phi_1) $ such that 
\ben
\phi_1 F(\Box) \phi_2 - \phi _2 F(\Box) \phi_2 = 
\p_\mu J^\mu (\phi_1,\phi_2) \,.  
\een
Thus on shell, 
\ben
\p_\mu J^\mu =0\,, 
\een 
and hence a conserved (pre)-symplectic form
on solutions $\phi_1$ and $\phi_2$ of (\ref{Pais}) is given by    
\ben
\omega( \phi_1,\phi_2) = - \omega( \phi_2,\phi_1)
= \int _\Sigma J^\mu d\Sigma_\mu \,,
\een
where $\Sigma$ is any Cauchy surface.  (The prefix ``pre'' would apply if
the symplectic form were degenerate.)

The equation of motion (\ref{Pais}) is clearly linear
and if  $F(\Box)$ is a polynomial of degree $N$, then 
of necessity quasi-linear,\footnote{In other words, all highest
derivative terms occur linearly.}  and hence its characteristic surfaces
$S={\rm constant}$ must, by standard theory,  satisfy 
\ben
(\eta ^{\mu \nu}\p_\mu S \p _\nu S  )^N  =0 \,.   
\een 
In other words the characteristic surfaces are null hypersurfaces.
Thus (\ref{Pais}) satisfies the \emph{Einstein-Causality
Principle.}

Since (\ref{Pais}) is linear we may substitute in it  the Ansatz
\ben
\phi = A e^{ik_\nu x^\mu }= A e^{i(-\omega t + \bk \cdot \bx)}  \,,   
\een 
where $A$ and $k_\mu$ are constant, and hence find that 
\ben
F(-k_\mu k ^\mu) =0 \,,
\een
where $k_\mu=(-\omega, {\bf k})$.
If the zeros of $F(u)$ are at $ u=m^2_i$ with real $m_i$, 
then we have several branches to the dispersion relation:
\ben
\omega = \pm \sqrt {m_i^2 + \bk ^2 } \,.
\een

\subsection{The Energy Momentum Tensor}

  In the simple case (\ref{non}), an energy-momentum tensor
is presented in \cite{Andrzejewski:2009bn}.  Making allowance for the 
conversion from the
$(+---)$ spacetime signature used in that reference, and 
also our choice for the overall sign for the Pais-Uhlenbeck action, 
this symmetric tensor is \cite{Andrzejewski:2009bn}
\bea
\bar T^{\mu \nu}&=& 2(\p^\mu \p^\nu \phi) \Box \phi - 
(\p^\mu \phi  ) \Box (\p ^\nu \phi) - (\p^\nu \phi) \Box (\p^\mu \phi)  
+(\p ^\mu \p ^\nu \p _\alpha \phi)( \p ^\alpha \phi) \nn\\
&& -
( \p ^\mu \p ^\alpha \phi ) (\p ^\nu \p _\alpha \phi ) 
-\half \eta ^{\mu \nu} (\Box \phi )^2 
-\half (m_1^2 + m_2 ^2 ) 
\bigl( \phi \p^\mu \p ^\nu \phi - \p^\mu \phi \p ^\nu \phi
 - \eta ^{\mu \nu}\phi \Box \phi  \bigr )\nn \\  
&&-  \half \eta ^{\mu \nu} m_1^2 m_2^2 \,\phi ^2 \,.\label{T1}
\eea
It obeys
\ben
\p_\mu \bar T^{\mu\nu} =  -(\p^\nu \phi) E \phi \,,
\een
and thus is conserved on shell.
Moreover if $\phi=\phi(t)$ then
\ben
\bar T^{00}= -\half \ddot \phi^2 + \dot \phi \, \dddot \phi  
+ \half (m_1^2 +m_2^2 ) \,\dot\phi^2 + \half  m_1^2 m_2^2\, \phi^2   
\label{T2}\,, 
\een
which agrees with the energy of the Pais-Uhlenbeck
oscillator derived using Noether's theorem. 

 The derivative terms in  (\ref{T1})  
differ from those in the standard energy-momentum tensor for a scalar field 
that one obtains by
the Belinfante-Rosenfeld \cite{Belinfante,Rosenfeld} procedure (i.e. 
 minimally coupling the action to a metric $g_{\mu \nu}$ 
and taking the variational derivative).  Later we shall
construct an energy-momentum tensor using the Belinfante-Rosenfeld procedure,
and in appendix A we show how it is related to (\ref{T1}).

\subsection{Second Quantisation of the Pais-Uhlenbeck Field Theory }

Recall that for a free theory we first construct 
the one-particle Hilbert space and
then construct the full (non-separable) Hilbert space using the
Fock construction.  The  standard approach is to complexify the Cauchy data   
and introduce a (symmetric)  sesquilinear form
\ben
h(\phi_1,\phi_2) =  -i \omega(\bar \phi_1, \phi_2) 
\een
where now $ \phi_1$ and $\phi_2$  are complex valued, and 
\ben \overline{
h(\phi_1,\phi_2 )} = h(\phi_2,\phi_1) \,.
\een

In the case of just two time derivatives 
in the equations of motion one has 
\ben
h(\phi_1,\phi_2) = -i \int_{{ E}^3}  \Bigl( 
\dot{\bar{\phi}}_1  \phi _2 - \bar{\phi}_1  
\dot \phi_2\Bigr) d^3 x \,.
\een
If $\phi_1=\phi_2 = Ae^{-i\omega t}$ 
we have 
\ben
h(\phi,\phi) = 2 \omega \int_{{ E}^3} |A |^2 d^3 x \,. 
\een 
Thus positive norm states have $\omega >0$ 
and so we identify the one-particle Hilbert space  
with the space of positive frequency Cauchy data.

\subsubsection{Symplectic Current for the Pais-Uhlenbeck Field Theory}

   When $m_1^2\ne m_2^2$, the general solution of the Pais-Uhlenbeck
equation $(\Box-m_1^2)(\Box-m_2^2)\phi=0$ is given by
\be
\phi= a \, \chi_1 + b\, \chi_2\,,
\ee
where $\chi_1$ and $\chi_2$ are solutions of $(\Box-m_1^2)\chi_1=0$ and
$(\Box-m_2^2)\chi_2=0$, and $a$ and $b$ are constants.  Thus $\chi_1$
and $\chi_2$ provide a basis for the general solution.  Without loss of 
generality, we shall assume $m_1^2 > m_2^2$ in what follows.

   We now write the field equation as 
\ben 
\Box^2 \phi  - (m_1^2 + m_2^2 ) \Box  \phi + m_1^2 m_2^2  \phi =0 \,.
\label{PUn}
\een
In Minkowski spacetime we have the following identities 
\bea
J^\mu _{\Box} &:=& \phi_1 \p ^\mu \phi_2 - ( \phi_1  \leftrightarrow\phi _2)  \,,\nn \\
\p_\mu J^\mu_{\Box}  &=& \phi_1 \Box \phi_2 - (\phi_1\leftrightarrow \phi_2 )
\,, \nn \\
J^\mu _{\Box ^2 }  &:=& \phi_1 \,\del^\mu \Box  \phi_2 +
    \Box\phi_1\, \del^\mu \phi_2 -(\phi_1 
\leftrightarrow \phi_2 ) \,,\nn  \\ 
\p_\mu   J^\mu _{\Box ^2 } &=& \phi_1 \Box^2 \phi_2
  - (\phi_1  \leftrightarrow \phi_2) \,. 
\label{currents} 
\eea
Thus if 
\ben
J^\mu  = J^\mu_{\Box^2} - (m_1^2 + m_2 ^2 ) J^\mu _{\Box} \,,
\een
then on shell
\ben
\p_\mu J^\mu =0\,.
\een

   Now suppose that $\phi_1$ and $\phi_2$ are any two solutions 
of
\ben
\Box \phi_1 = \mu_1^2 \phi_1 \,, \qquad \Box \phi_2 = \mu^2_2 \phi_2 \,,
\een
where each of $\mu_1^2$ and $\mu_2^2$ can be either $m_1^2$ or
$m_2^2$.
Then 
\bea
J^\mu &=& \Bigl ( \mu_1^2 + \mu_2^2)  - (m_1^2 +  m_2^2)  \Bigr ) 
(\phi_1 \p^\mu \phi_2 - \phi_ 2 \p^\mu \phi_1) \\    
&=&  \Big( ( \mu_1^2 + \mu_2^2)  - (m_1^2 + m_2^2)  \Big)  J^\mu _\Box \,.
\eea
Evidently if $\phi_1$ and $\phi_2$ have different masses (i.e. if
$\mu_1^2=m_1^2$ and $\mu_2^2=m_2^2$, or else if $\mu_1^2=m_2^2$ and
$\mu_2^2=m_1^2$), then
$J^\mu =0$. 

If $\phi_1$ and $\phi_2$ have equal mass-squared  $\mu_1^2=\mu_2^2=m_1^2$ then
\ben
J^\mu = (m_1^2 -m_2^2) J^\mu _\Box \,.
\een    
while if  $\phi_i$ and $\phi_2$ have equal mass-squared 
$\mu_1^2=\mu_2^2= m_2^2$ then 
 \ben
J^\mu = (m_2^2 -m_1^2) J^\mu _\Box \,.
\een 
If we choose the complex modes with mass $m_1$ to have positive frequency 
and the complex modes with mass $m_2$ to to have negative
frequency, the Hermitian  form $h(\cdot\, , \cdot)$ 
will be positive definite and diagonalised in this basis.

Thus, despite the fact that we associate 
one creation operator with a positive frequency
mode 
and the other with a negative frequency mode,   
we obtain  a positive norm on our Hilbert space.
There are no states with negative norm (sometimes called ``ghosts''). 
Of course if $m_2^2 > m_1^2$ we must reverse the convention.

\section{Pais-Uhlenbeck Field Theory in Curved spacetime}

\subsection{Simplest model}

We now replace 
$\p_\mu$ by $\nabla _\mu$, the Levi-Civita covariant derivative, and so
$\Box= \nabla _\mu \nabla ^\mu$,
the generally-covariant d'Alembertian. 
The field equation is
\ben
E \phi \equiv \bigl( \Box^2 - (m_1^2 + m_2^2 ) \Box  + 
m_1^2 m_2 ^2 ) \bigr ) \phi =0\,. \label{EOM}   
\een
The field equation may be derived from the action
\be
S = -\ft12 \int \sqrt{-g}\, d^4x\, \Big[(\Box\,\phi)^2 +
(m_1^2 +m_2^2)\, g^{\mu\nu}\, \del_\mu\phi\, \del_\nu\phi +
m_1^2 m_2^2\, \phi^2\Big]\,,\label{action0}
\ee
from which one finds that under a variation of $\phi$, 
\bea
\delta S=  - \int _D \sqrt{-g} d^4x\, \delta \phi E \phi
+\int _{\p D}  \Bigl( \nabla^\mu \Box\,\phi\, \delta\phi
- \Box\, \phi\, \nabla^\mu\delta\phi  -(m_1^2 +m_2^2 )
\nabla^\mu \phi\,\delta\phi  \Bigr ) d \Sigma _\mu  \,.
\eea

The operator $E$ may also be written as 
\ben 
E= (\Box- m_1^2)(\Box - m_2^2 )= 
(\Box - m_2^2)(\Box - m_1^2 )\,. 
 \een
The identities
\bea
\chi \Box\, \phi &=& \nabla _\mu (\chi \nabla ^\mu \phi) - g_{\mu \nu} \nabla
^\mu \chi \nabla ^\nu \phi\,, \\ 
\chi \Box^2 \,\phi &=& \nabla _\mu
\Bigl ( \chi \nabla^\mu\Box\, \phi -  (\nabla^\mu \chi) \Box\, \phi \Bigr )  
+ (\Box \,\chi)(\Box\, \phi) \,, \label{ids} 
\eea
imply  the formal self-adjointness of $E$ 
since,  from them,   one deduces that 
\ben
\chi E \phi-\phi E \chi = \nabla_\mu  J^\mu \,, 
\een
where 
\ben
J^\mu[\chi,\phi] = \chi \nabla^\mu \Box\, \phi - (\nabla^\mu \chi)
\Box\, \phi
- (m_1^2 + m_2^2 )\chi \nabla ^\mu \phi - (\chi \leftrightarrow \phi) \,.  
\een
If $\chi$ and $\phi$ are two solutions of the equation of motion (\ref{EOM}) 
then one deduces that
\ben
\nabla _\mu J^\mu[\chi,\phi] =0 \,, 
\een  
and hence that the (pre-)symplectic form $\omega(\chi,\phi)$ 
on the space of solutions of (\ref{EOM}),  given by
\ben
\omega (\chi,\phi) = \int _{\Sigma} J^\mu[\chi,\phi] \, d \Sigma_\mu
= - \omega(\phi,\chi) \,,
\een
where $\Sigma$ is a Cauchy surface, is conserved.

Now we suppose that $\chi=\phi_1$ satisfies $\Box\, \phi_1 = \mu_1^2\, \phi_1$
and $\phi=\phi_2$ satisfies $\Box\,\phi_2 = \mu_2^2\, \phi_2$,
where each of $\mu_1^2$ and $\mu_2^2$ can each be either  $m_1^2$ or  $m_2^2$.  
Then 
\ben
J^\mu =(\mu_1^2 +\mu_2^2 - m_1^2 - m_2 ^2 ) 
( \phi_1 \nabla ^\mu \phi_2 - \phi_2 \nabla ^\mu \phi_1 ) \,.
\een    
Thus the space of Cauchy data and the symplectic form splits as a direct sum 
of two summands, labelled by the mass-squared $m_1^2 $ and $ m_2^2 $. 
    
\subsubsection{Auxiliary field formulation}  

   A convenient way to handle the higher-derivative terms in the
Pais-Uhlenbeck equation is by introducing auxiliary fields.  In particular,
as we shall see below, this provides a simple way to calculate the
contributions to the energy-momentum tensor from the higher-derivative
terms.  We shall employ two related, but slightly different, approaches
in this paper.  For now, we shall handle the various higher-derivative
terms that arise in a general Pais-Uhlenbeck theory separately, by
developing an auxiliary-field construction for a Lagrangian
${\cal L}_n = -\ft12 \sqrt{-g}\, \phi\,(-\Box)^n\,\phi$.  The
general Pais-Uhlenbeck Lagrangian is then given as a linear combination
of such terms with certain constant coefficients.  The energy-momentum
tensor for the PU theory is then similarly given by the analogous
linear combination of the associated energy-momentum tensors.  In appendix
B, we discuss a slightly different approach, in which the Lagrangian for
the full PU field theory is directly rewritten by introducing auxiliary
fields.

   In the present simplest example we are considering, for which
the PU action is given by 
(\ref{action0}), there is just the one higher-derivative term, namely
\ben  
S_\2 [\phi] = -\half \int _D \sqrt{-g}  d^4 x\,     
  (\Box\, \phi)^2 \,.\label{action1}
\een
We now introduce an auxiliary field $\psi$ and replace (\ref{action1}) 
by 
\ben
\widetilde S_\2 [\phi,\psi] = 
\int _D \sqrt{-g}  d^4 x\, \Bigl ( -\psi \Box \phi + \half \psi ^2 \bigr )
\label{action2} \,. 
\een
The variation of (\ref{action2}) with respect to $\psi$ yields 
\ben
\psi = \Box\, \phi\,, 
\label{psieqn} 
\een  
and upon substituting this back into (\ref{action2}) one indeed recovers 
(\ref{action1}).
The variation of (\ref{action2}) with respect to $\phi$ gives 
\ben
\Box\, \psi =0\,, 
\een   
whence  we obtain the field equation $\Box^2\,\phi=0$ for  $\phi $, 
which is indeed the same as the equation that arises directly from 
(\ref{action1}).

\subsubsection{Energy-momentum tensor} 

Up to an integration by parts the action (\ref{action2}) is 
equivalent to
\ben
\widetilde S_\2[\psi,\phi, g^{\mu \nu} ] = \int _D \sqrt{-g}  d^4 x\, 
\Bigl (\p_\mu \psi \, 
 g^{\mu \nu} \p_\nu \phi
+ \half \psi ^2 \Bigr )\,.
\een
 Now the associated Belinfante-Rosenfeld 
energy-momentum tensor $T^\2_{\mu \nu}$ is given by   
\bea
T^\2_{\mu \nu} &=& -2 
 \frac{\delta}{\delta g^{\mu \nu}}  \tilde S_2[\psi,\phi,g^{\mu \nu}]\\
&=&  -\p_\mu \psi \p_\nu \phi - \p_\nu \psi \p_\mu \phi + g_{\mu \nu }
\p_\alpha \psi g^{\alpha \beta} \p _\beta \phi +\half g_{\mu \nu} \psi ^2 \,.   
\label{aux}
\eea
Substituting (\ref{psieqn}) in (\ref{aux}) therefore gives
\ben 
T^\2_{\mu \nu} = -\p_\mu \Box\,\phi\,  \p_\nu \phi - 
\p_\nu \Box\, \phi\, \p_\mu \phi  + g_{\mu \nu} \p_\alpha \Box\, \phi\,
\p^\alpha \phi + \half g_{\mu \nu} \,(\Box\,\phi)^2 \,,\label{4thorderT} 
\een 
as the contribution to the energy-momentum tensor of the Pais-Uhlenbeck
theory from the $(\Box\,\phi)^2$ term in (\ref{action0}).
One may verify that
\ben
\nabla^\mu T^\2_{\mu \nu} = -\Box^2\, \phi \, \p_\nu \phi \,,
\een
and so  indeed this contribution to 
$T_{\mu \nu}$ is covariantly conserved if the field equation
$\Box^2\,\phi=0$ holds.

  Combining the contribution $T_{\mu\nu}^\2$ with the contributions
from the remaining terms in (\ref{action0}), which are straightforward
to calculate in the standard way, we find that the total energy-momentum
tensor for this PU theory is given by
\bea
T_{\mu \nu} &=& -\p_\mu \Box\,\phi\,  \p_\nu \phi -
\p_\nu \Box\, \phi\, \p_\mu \phi  + g_{\mu \nu} \p_\alpha \Box\, \phi\,
\p^\alpha \phi + \half g_{\mu \nu} \,(\Box\,\phi)^2 \nn\\
&& +(m_1^2 + m_2 ^2 ) \big( \p_\mu \phi \p_\nu \phi - \half g_{\mu \nu}
g^{\alpha \beta} \p_\alpha \phi \p_\beta \phi\big) -
\half g_{\mu \nu} \,m_1^2 \,m_2^2 \,\phi ^2 \,.  \label{PU4T} 
\eea   

However it should be noted that  $T_{\mu \nu}$
is different, in its flat-spacetime limit, 
from the expression for $\bar T_{\mu\nu}$ given by (\ref{T1}). 
In appendix A we discuss a general derivation of the canonical energy-momentum
tensor derived by the Noether method in flat spacetime, and its relation
to the energy-momentum tensor that we obtained in (\ref{PU4T}) by
using the Belinfante-Rosenfeld procedure.

Interestingly, although the flat space limit of 
(\ref{PU4T}) is not the same as (\ref{T1}), it gives the same
expression (\ref{T2}) if one specialises to 
$\phi=\phi(t)$.

\subsection{Auxiliary Fields}\label{auxfieldsec}

\subsubsection{Powers of Laplace/d'Alembert operator}

Consider now the $n$'th power of the d'Alembert operator $\Box=\nabla^2$. 
By introducing auxiliary fields $\chi_k$ and
Lagrange multipliers $\eta_k$, for $k=1,\ldots,n-1$, 
we can re-write
\bea
\phi\Box^n\phi&\rightarrow&\phi\Box\chi_1+\eta_1(\Box \chi_2-\chi_1)
+\eta_2(\Box \chi_3-\chi_2)+ \cdots \nn\\
&& +
 \eta_{n-2}(\Box \chi_{n-1}-\chi_{n-2})+\eta_{n-1}(\Box\phi -\chi_{n-1})\,.
\lb{7}
\eea
Indeed, varying with $\eta_k$ we find
\be
\chi_1=\Box \chi_2\, , \ \ \chi_2=\Box\chi_3\, , \ \ \cdots \chi_{n-1}=
\Box\phi \Rightarrow  \chi_1=\Box^{n-1}\phi\,.
\lb{8}
\ee
On the other hand, varying with respect to $\chi_k$ we find that
\be
\eta_1=\Box \phi\, , \ \ \eta_2=\Box \eta_1\, , 
\ \ \cdots \ \ \eta_{n-1}=\Box^{n-2}\phi\,.
\lb{9}
\ee
This shows that we may identify 
\be
\eta_k=\chi_{n-k}\, , \ \ k=1,\ldots, n-1\,. 
\label{etachi}
\ee
This identification reduces the number of extra fields by 2. 
We have, for instance,
\be
&&\phi \Box^2\phi = \phi\Box\chi+\chi\Box\phi-\chi^2 \,,\nonumber \\
&&\phi\Box^3\phi=\phi\Box \chi_1+\chi_2\Box\chi_2+\chi_1\Box 
\phi-2\chi_2\chi_1 \lb{11}\,.
\ee
   If we also define $\chi_0\equiv0$ and $\chi_n\equiv\phi$, then after 
making the identifications
(\ref{etachi}) we see that 
\be
\phi\,\Box^n\,\phi  = 
   \sum_{k=0}^{n-1}\chi_{n-k}(\Box \chi_{k+1}-\chi_k)\,.
\label{chilag}
\ee
Note that combined with the solutions (\ref{etachi}) for the auxiliary fields,
we have
\be
\chi_k= \Box^{n-k}\, \chi\,,\qquad 1\le k\le n\,,\qquad \chi_0=0\,.
\ee

   Consider the Lagrangian
\be
{\cal L} = \sqrt{-g}\, \Big[-\ft12  \phi\, (-\Box)^n\, \phi - V\Big]\,,
\ee
whose equations of motion are
\be
(-\Box)^n\, \phi = -\fft{\del V}{\del\phi}\,.
\ee
As established above, this can be written, after integrations by parts, as
\be
{\cal L} = \ft12 (-1)^n\, \sqrt{-g}\, \sum_{k=0}^{n-1} 
\Big[ \del_\mu\chi_{n-k}\, \del^\mu \chi_{k+1} + \chi_{n-k}\, \chi_k\Big]
-\sqrt{-g}\, V\,,
\ee
with the auxiliary fields given by
\be
\chi_k = \Box^{n-k}\, \phi\,,\qquad 1\le k \le n\,,\qquad \chi_0=0\,.
\label{chis}
\ee

   Calculating the energy-momentum tensor $T_{\mu\nu}= -(2/\sqrt{-g})\, 
\delta{\cal L}/\delta g^{\mu\nu}$, we thus find
\bea
T_{\mu\nu} &=& (-1)^n \sum_{k=0}^{n-1} \Big[ -\del_\mu \chi_{n-k}\,
    \del_\nu \chi_{k+1}  +
     \ft12 g_{\mu\nu}\, \del_\rho\chi_{n-k} \, \del^\rho \chi_{k+1}
\Big] \nn\\
&&  +\ft12 (-1)^n\, g_{\mu\nu}\,\sum_{k=1}^{n-1}
  \chi_{n-k} \, \chi_k - g_{\mu\nu}\, V\,,\label{Tmunuchi}
\eea
and hence, using (\ref{chis}),
\bea
T_{\mu\nu} &=& (-1)^n \sum_{k=0}^{n-1} \Big[ -\del_\mu \Box^k\phi\, 
    \del_\nu \Box^{n-k-1}\phi + 
     \ft12 g_{\mu\nu}\, \del_\rho \Box^k\phi\, \del^\rho \Box^{n-k-1}\phi
\Big] \nn\\
&&  +\ft12 (-1)^n\, g_{\mu\nu}\,\sum_{k=1}^{n-1}  
  \Box^k\phi\, \Box^{n-k}\phi  - g_{\mu\nu}\, V\,.\label{Tmunu}
\eea
(Note that the first term is in fact 
automatically symmetric in $\mu$ and $\nu$, in view of 
the summation over $k$.)
The first few examples, for $n=1$, 2 and 3, are:
\bea
n=1:&& T_{\mu\nu}= \del_\mu\phi\, \del_\nu\phi - \ft12 g_{\mu\nu}\,
  \del_\rho\phi\, \del^\rho\phi - g_{\mu\nu}\, V\,,\nn\\
n=2:&& T_{\mu\nu}= -\del_\mu\phi\, \del_\nu\Box\phi 
          -\del_\nu\phi\, \del_\mu\Box\phi+ 
   g_{\mu\nu}\, \del_\rho\Box\phi\,\del^\rho\phi +
   \ft12 g_{\mu\nu}\, (\Box\phi)^2 - g_{\mu\nu}\, V\,,\nn\\
n=3:&& T_{\mu\nu}= \del_\mu\phi\, \del_\nu\Box^2\phi +
   \del_\mu\Box\phi\, \del_\nu\Box\phi +
    \del_\mu\Box^2\phi\, \del_\nu\phi  -g_{\mu\nu}\, \del_\rho\Box^2\phi\,
   \del^\rho\phi - \ft12 g_{\mu\nu}\, \del_\rho\Box\phi\, \del^\rho\Box\phi
   \nn\\
&& \qquad\quad 
   - g_{\mu\nu}\, \Box\phi\, \Box^2\phi - g_{\mu\nu}\, V\,.
\eea

\subsubsection{Auxiliary Fields for General Higher-Order Operators}

  Consider the higher-derivative Lagrangian 
\be
L= \phi\, A_1 A_2\cdots A_n\, \phi\,,\label{nonaux}
\ee
where the operators $A_i$ are formally self-adjoint 
and need not necessarily be mutually commuting.  
An example for which they are mutually commuting is 
\be
A_i= \Box-  m_i^2\,,\qquad \Box = \nabla^2\,.\label{Aimass}
\ee
Paralleling the procedure for introducing auxiliary fields for the 
Lagrangian $L_n = \phi\Delta^n\phi$, consider here
\bea
L&=& \phi A_1\, \chi_1 + \eta_1(A_2\, \chi_2 - \chi_1) + 
  \eta_2(A_3\, \chi_3 - \chi_2) + \cdots \nn\\ 
&+&
 \eta_{n-2}\, (A_{n-1}\, \chi_{n-1} -\chi_{n-2}) +
\eta_{n-1} (A_n\, \phi - \chi_{n-1})\,.\label{auxform}
\eea
Varying the $\eta_i$ fields gives
\bea
\chi_1 &=& A_2\, \chi_2\,,\qquad \chi_2= A_3\, \chi_3\,, \qquad\cdots , \qquad 
 \chi_{n-1}= A_n\, \phi\,,\nn\\
&\implies & \chi_1 = A_2 A_3 \cdots A_n\, \phi\,.\label{chisol}
\eea
Varying the $\chi_i$ fields gives
\be
\eta_1=A_1\, \phi\,,\qquad \eta_2= A_2\, \eta_1\,,\qquad \cdots ,\qquad 
   \eta_{n-1}= A_{n-1}\, \eta_{n-2}\,.\label{etasol}
\ee
Unlike in the case of the pure $\Box^n$ Lagrangian, we cannot equate
the set of $\eta_i$ fields with the set of $\chi_i$ fields.

  It is easily seen that the Lagrangian (\ref{auxform}) gives rise to 
(\ref{nonaux}) after plugging in the solutions for the $\chi_i$ fields,
given in (\ref{chisol}).  Note that if we define
\be
\chi_0\equiv0\,,\qquad \chi_n\equiv \phi\,,\qquad \eta_0\equiv\phi\,,\qquad
  \eta_n\equiv 0\,,
\ee
then the Lagrangian (\ref{auxform}) can be written as
\be
L= \sum_{k=0}^{n-1} \eta_k\, (A_{k+1}\, \chi_{k+1} - \chi_k)\,.\label{genlag}
\ee
Note that together with the solutions (\ref{chisol}) and (\ref{etasol}) for
the auxiliary fields, we have\footnote{The discussion we gave in 
section \ref{auxfieldsec}, where we introduced auxiliary fields for the 
Lagrangian $\phi\, \Box^n\,\phi$, is a special case of the our discussion 
here, in which $A_k=\Box$ for all $k$.  As can be seen from (\ref{chisol}) and
(\ref{etasol}), in this special case one can equate $\eta_k=\chi_{n-k}$
as in section \ref{auxfieldsec}.  But in the more general case discussed in 
this section, one cannot equate the $\eta$ and $\chi$ fields, even in
the case that the operators $A_i$ commute.  If, however, the operators
$A_i$ take the form given in (\ref{Aimass}), where the
differential operator parts of all the $A_i$ are the same, then there
exist purely algebraic relations between the set of
$\eta_k$ fields and the set of $\chi_k$ fields,
as can be seen from (\ref{chieta}). In this case, therefore, one
needs only $\phi$ and either the set of $\chi_k$ or the set of $\eta_k$
as auxiliary fields.}
\bea
\chi_k &=& A_{k+1}\, A_{k+2}\, \cdots A_n\,\phi\,,\qquad 1\le k\le n\,,\qquad
  \chi_0=0\,,\nn\\
\eta_k&=& A_k\, A_{k-1}\cdots A_1\, \phi\,,\qquad 0\le k \le n-1\,,\qquad
\eta_n=0\,.\label{chieta}
\eea

  If we consider the example where $A_i$ is given by (\ref{Aimass}),
the energy-momentum tensor that follows from (\ref{genlag}) is
\be
T_{\mu\nu}= \sum_{k=0}^{n-1} \Big[ \del_{(\mu} \eta_k\, \del_{\nu)} \chi_{k+1}
  -\ft12 g_{\mu\nu}\, \Big( \del_\rho \eta_k\, \del^\rho \chi_{k+1} 
+ m_{k+1}^2\, \eta_k\, \chi_{k+1} + \eta_k\, \chi_k\Big)\Big]\,.
\ee

\subsection{Symplectic Current in Higher-Derivative Scalar Theory}

  Consider first the Lagrangian $\half \phi\square^n\phi$, whose equation of
motion is
\be
\square^n\phi=0\,.
\ee
By extracting derivatives in sequence from
\be
\psi\, \square^n\, \phi-\phi\,\square^n\, \psi\,, 
\ee
we obtain
\bea
\psi\,\square^n\, \phi-\phi\,\square^n\, \psi 
&=& \nabla_\mu\Big(\psi\, \nabla^\mu\square^{n-1}\, \phi
\Big) -\nabla_\mu\psi\, \nabla^\mu\square^{n-1}\, \phi 
-(\psi\leftrightarrow\phi)\,,\nn\\
&=&\nabla_\mu\Big(\psi\, \nabla^\mu\square^{n-1}\, \phi-
             \nabla^\mu\psi\, \square^{n-1}\, \phi \Big) +
   \square\psi\, \square^{n-1}\phi 
-(\psi\leftrightarrow\phi)\,,\nn\\
&=& \cdots\,,\nn\\
&=& \sum_{p=0}^{n-1}  J^\mu(\square^p\,\psi,\square^{n-p-1}\, \psi)
\,,
\eea
where we have defined
\be
 J^\mu(A,B) \equiv A\nabla^\mu\, B - B \nabla^\mu\, A\,.
\ee
Thus, if $\phi$ and $\psi$ are any solutions of $\square^n\phi=0$ and
$\square^n\psi=0$, we have a conserved current
\be
J^\mu_{\square^n}(\psi,\phi) = 
\sum_{p=0}^{n-1}  J^\mu(\square^p\,\psi,\square^{n-p-1}\, \phi) 
\,.\label{Jsqn}
\ee

 Consider now the Lagrangian, and consequent equations of motion,
\bea
{\cal L} &=& \half \phi\, \prod_{i=1}^N (\square-m_i^2)\, \phi\,,\nn\\
\Delta_N\,\phi &\equiv& \prod_{i=1}^N (\square-m_i^2) \, \phi=0\,.
\label{DelNdef}
\eea
If we assume the masses are all unequal, the general solution is a linear
combination of ``elementary solutions'' satisfying
\be
(\square-m_i^2)\, \phi_i=0\,,\label{elementary}
\ee
for any $i$ in the range $1\le i\le N$.

   We now consider the conserved current $J_{\Delta_N}^\mu$ 
for the 
operator $\Delta_N$ defined in (\ref{DelNdef}).  If we write the
operator as
\be
\Delta_N = \sum_{n=0}^N a_n\, \square^n\,,
\ee
then this current is given by
\be
J_{\Delta_N}^\mu(\psi,\phi) = \sum_{n=1}^N a_n\, 
J^\mu_{\square^n}(\psi,\phi)\,,
\ee
where $J^\mu_{\square^n}(\psi,\phi)$ is defined in (\ref{Jsqn}).  

   There are two cases of particular interest to consider.  The first is
when $\psi$ and $\phi$ are two ``elementary'' solutions of $\Delta_N$, 
as defined in (\ref{elementary}), 
corresponding to two different $m_i^2$ values.  Without loss of generality
we may take $\psi=\phi_1$ and $\phi=\phi_2$, that is to say they satisfy
\be
\square\psi=m_1^2\, \psi\,,\qquad \square\,\phi= m_2^2\, \phi\,.
\ee
Writing the operator $\Delta_N$ as 
\be
\Delta_N= (\square-m_1^2)(\square-m_2^2)\, f(\square)\,,
\ee
with
\be
f(\square)= \prod_{i=3}^N (\square-m_i^2)= \sum_{k=0}^{N-2}\, b_k\, \square^k
\,,
\ee
and noting that
\be
J^\mu(\square^p\,\psi,\square^{(n-p-1)}\, \phi)=
   m_1^{2p}\, m_2^{2n-2p-2}\, J^\mu_\square(\psi,\phi)\,,
\ee
we see that
\bea
J^\mu_{\Delta_N}(\psi,\phi) &=& 
\sum_{k=0}^{N-2} b_k\, \Big[\sum_{p=0}^{k+1} m_1^{2p}\, m_2^{2k-2p+2} 
  -(m_1^2+m_2^2)\, \sum_{p=0}^k m_1^{2p}\, m_2^{2k-2p} \nn \\
&& \qquad\qquad
   +
  m_1^2\, m_2^2\, \sum_{p=0}^{k-1}\, 
   m_1^{2p}\, m_2^{2k-2p-2}\Big]\, J^\mu_\square(\psi,\phi)\,.
\eea
Collecting the terms in the square brackets as a sum $\sum_{p=0}^{k-1}$
together with the additional terms from the first two summations we 
immediately find that the total vanishes for each $k$.  Thus we
find
\be
J^\mu_{\Delta_N}(\psi,\phi) =0\,,
\ee
whenever $\psi$ and $\phi$ are elementary solutions with different mass
values.

   The other case of interest is when $\psi$ and $\phi$ are solutions
with the same mass value.  Without loss of generality, we may consider the
case where $\psi$ and $\phi$ satisfy
\be
\square\,\psi=m_1^2\, \psi\,,\qquad \square\,\phi=m_1^2\, \phi\,.
\label{equalmass}
\ee
This means that $J^\mu_{\square^n}(\psi,\phi)= 
m_1^{2n-2}\, J^\mu_\square(\psi,\phi)$.  Writing the operator $\Delta_N$
as
\be
\Delta_N =(\square-m_1^2)\, h(\square)
\ee
with
\be
h(\square)= \prod_{i=2}^N (\square-m_i^2)= \sum_{k=0}^{N-1}  c_k\, \square^k\,,
\ee
we find that
\bea
J^\mu_{\Delta_N}(\psi,\phi) &=& 
\sum_{k=0}^{N-1} c_k\, \Big[ J^\mu_{\square^{k+1}}(\psi,\phi)-m_1^2\,
   J^\mu_{\square^k}(\psi,\phi)\Big]\,,\nn\\
&=& \sum_{k=0}^{N-1} c_k\,
  \Big[ (k+1)\, m_1^{2k} - k\, m_1^{2k}\Big] \, J^\mu_\square(\psi,\phi)\,,\nn\\
&=& \sum_{k=0}^{N-1} c_k\, m_1^{2k}\, J^\mu_\square(\psi,\phi)\,,\nn\\
&=& h(m_1^2)\, J^\mu_\square(\psi,\phi)\,,\nn\\
&=& \prod_{i=2}^{N} (m_1^2-m_i^2)\, J^\mu_\square(\psi,\phi)\,.
\eea
In other words, when $\psi$ and $\phi$ are solutions
with the same mass, as in  (\ref{equalmass}), the conserved current is 
given by
\be
J^\mu_{\Delta_N}(\psi,\phi)= \prod_{i=2}^{N} (m_1^2-m_i^2)\, J^\mu(\psi,\phi)
\,,
\ee
where $J^\mu(\psi,\phi)$ is the usual current
\be
J^\mu(\psi,\phi)= \psi\, \nabla^\mu\, \phi- \phi\,\nabla^\mu\, \psi\,.
\ee

In the general case when $\square\psi=y^2\psi$ and $\square\phi=x^2\phi$,
we have 
\be
\Delta_N= \prod_{i=1}^N (\square-m_i^2) = W(\square)=\sum_{k=0}^N a_k\, 
 \square^k\,,
\ee
and 
\be
J^\mu_{\Delta_N}(\psi,\phi)= \sum_{k=1}^N a_k\, 
              J^\mu_{\square^k}(\psi,\phi)\,.
\ee
 From (\ref{Jsqn}) we therefore have 
\bea
J^\mu_{\Delta_N}(\psi,\phi) &=&  \sum_{k=1}^N a_k\,x^{2k-2}\, \sum_{p=0}^{k-1}  
  \Big(\fft{y}{x}\Big)^{2p}  \, J^\mu(\psi,\phi)\,,\nn\\
&=& \sum_{k=1}^N a_k\, x^{2k-2}\,  \Big(\fft{1-\Big(\fft{y}{x}\Big)^{2k}}{
    1-\Big(\fft{y}{x}\Big)^2}\Big)\, J^\mu(\psi,\phi)\,,\nn\\
&=& \fft1{x^2-y^2}\, \sum_{k=1}^N a_k\, (x^{2k}-y^{2k})\, J^\mu(\psi,\phi)
\,,\nn\\
&=& \fft1{x^2-y^2}\, \sum_{k=0}^N a_k\, 
  (x^{2k}-y^{2k})\, J^\mu(\psi,\phi)\,,\nn\\
&=& \fft{1}{x^2-y^2}\, (W(x^2)-W(y^2))\, J^\mu(\psi,\phi)\,,
\eea
and hence
 we have 
\be
J^\mu_{\Delta_N}(\psi,\phi)=\frac{\prod_{i=1}^N(y^2-m_i^2)-
    \prod_{i=1}^N(x^2-m_i^2)}{y^2-x^2}\, J^\mu(\psi,\phi)\,.
\ee

\section{Green Functions and Euclidean Formulation}

Pais and Uhlenbeck solved for
the time-independent Green function $G^{(4)}(\br)$ for the 4th-order
operator in flat space, defined by  
\ben
(-\nabla ^2 +m_1^2 ) (-\nabla ^2 +m_ 2 ^2)  G^{(4)}(\br) = \delta(\br)\,,
\label{Green} 
\een 
where $\br=\br_1-\br_2$, by using Fourier transforms.
Since
\ben
\frac{1}{ (-\nabla ^2 +m_1^2 ) (-\nabla ^2 +m_ 2 ^2)} = 
\frac{1}{m_2^2 - m_1 ^2} \Bigl[\frac{1}{ (-\nabla^2  + m_1^2) } -  
\frac{1}{(-\nabla ^2  + m_2^2)}     \Bigr ] \,,
\een
the formal solution\footnote{Pais and Uhlenbeck omit the denominator
 $(m_2^2 - m_1^2)$.}   is 
\ben
G^{(4)}(\br) = 
\fft{\bigl( e^{-m_1r } - e^{-m_2r}  \bigr )}{4\pi (m_2^2-m_1^2)\,r}   
\,,\label{greenf}
\een
where $r=|\br|$. $G^{(4)}(\br)$ goes to a constant value at the origin,
and it admits a Taylor expansion in powers of $r$.  However,
because the expansion includes odd powers of $r$ it is not differentiable
with respect to the Cartesian coordinates at the origin $r=0$, and 
indeed for this reason it
does in fact satisfy (\ref{Green}) with the delta function on the right-hand
side.  This may be easily verified by integrating (\ref{Green}) over a small 
spherical volume centred on the origin, and using the divergence theorem
to turn the derivative terms into surface integrals.
Note that $G^{(4)}(\br)$ is symmetric 
under the interchange of $m_1$ and $m_2$.
Both $(-\nabla ^2 + m_1^2)$ and $(-\nabla ^2 + m_2^2)$
are positive operators, and  their  Green 
functions $\frac{e^{-m_1 r}}{4 \pi r}$ and   $\frac{e^{-m_2 r}}{4 \pi r}$ 
are both positive. The operator 
$ (-\nabla ^2 +m_1^2) (-\nabla ^2 + m_ 2 ^2)$ is positive
and, as one may check, $G^{(4)}(\br)$ is also positive.

The Green function (\ref{greenf}) is unique, 
since there are no non-singular solutions of the homogeneous equation  
\ben 
\nabla ^4 \phi  -( m_1^2+ m_2^2 ) \nabla ^2 \phi + m_1^2 m_1^2 \phi =0 
\label{homog}
\een
that fall off sufficiently fast at infinity.  This is easily 
established by multiplying (\ref{homog}) by $\phi$ and integrating over 
${ E} ^3$. Integrating by parts  gives
\ben
\int _{{E} ^3}  \Bigl[ (\nabla^2  \phi )^2 
+ (m_1^2 + m_2^2 )( \nabla \phi )^2 + m_1^2\, m_2^2\, \phi^2 \Bigr] d^3 x  =0\,,
\een
whence $\phi=0$. 

Similar conclusions hold in four dimensional Euclidean space,
but the explicit form of the Green
function is more complicated since it involves Bessel functions.  Concretely,
the Euclidean Green function $G_E(r)$ for the 2nd-order Klein-Gordon
operator, obeying  
$(-\nabla^2 + m^2) G_E(r)=\delta(\br)$, is given by
\be
G_E(r) = \fft{m\, K_1(mr)}{4\pi^2\, r}\,,\label{Geuc}
\ee
where $K_\nu(x)$ is the modified Bessel function of the second kind,
which has the integral representation
\be
K_\nu(x) = \fft{\sqrt{\pi}\, x^\nu}{2^\nu\, \Gamma(\nu+\ft12)}\,
\int_1^\infty dt\, e^{-xt}\, (t^2-1)^{\nu-\ft12}\,.\label{Knuint}
\ee
At small $r$ the Green function $G_E(\br)$ has the expansion
\be
G_E(\br) = \fft1{4 \pi^2\, r^2} + \fft{m^2}{16\pi^2}\, \Big(2\gamma +
   2 \log\fft{m r}{2}-1\Big)+\cdots\,,\label{greenE2}
\ee
where $\gamma$ is the Euler-Mascheroni constant.  For the 4th-order
Pais-Uhlenbeck operator, the corresponding Green function $G_E^{(4)}(r)$
obeys (\ref{Green}) (with $\nabla^2$ now understood to be the four-dimensional
Euclidean Laplacian), and is given by
\be
G_E^{(4)}(\br)=
 \fft{m_1\, K_1(m_1 \,r)- m_2 \, K_1(m_2\, r)}{4\pi^2\, (m_2^2-m_1^2)\, r}
\,.\label{greenE4}
\ee
One may easily verify, for example by using (\ref{Knuint}),
that both (\ref{greenE2}) and (\ref{greenE4}) are positive.
(Note that the $r^{-2}$ singularity in (\ref{greenE2}) is cancelled in
(\ref{greenE4}).)
Because of the positivity of $G_E^{(4)}(\br)$, one 
might be led to believe that the Osterwalder-Schrader construction
of the quantum field theory could be carried  out with no
negative norm states. This however is not the case \cite{Hawking:1985gh}, 
as we shall discuss below.

\subsection{Reflection Positivity in the 4th-Order Case}

If we work on a Riemannian manifold 
the Euclidean action functional  
\ben
I= \half \int _D \sqrt{g}\, d^4x\,  \Bigl[ 
  (\nabla ^2 \phi)^2 + (m_1^2 + m_2^2 ) 
( g ^ {\mu \nu}  \p _\mu \phi \p _\nu   \phi) 
+ m_1^2\, m_2^2\, \phi ^2       \Bigr ] 
\een
is positive definite. Thus the Euclidean functional integrals should make sense
and the associated two-point function or Green function
\ben
G_E  = \frac{1}{m_2^2 - m_1^2} \Bigl 
[ \frac{1}{(-\nabla ^2 + m_1^2) } - \frac{1}{(-\nabla ^2 + m_2^2) }  \bigr]\,,  
\een
being the inverse of the positive operator, will be positive.

In general, given a Riemannian manifold there is no Lorentzian
manifold associated with it by analytic continuation.
If there is, one may analytically continue the two-point function and ask
what, if any, is its physical significance.
In the case of Euclidean space $\mathbb{E}^4$ it is well known that one
obtains the Wightman function associated
with the vacuum state. If one periodically identifies one of the
coordinates with period $\beta$, one obtains expectation values in
the Gibbs state at temperature $T=\frac{1}{\beta}$. In fact using
a construction due to Osterwalder and Schrader one need
not depart from Euclidean space in order to construct the
quantum mechanical Hilbert space. This will be possible as
long as the Green function satisfies the requirement of
Reflection Positivity \cite{Osterwalder:1973dx}.\footnote{For a helpful 
introduction the reader
may consult \cite{Uhlmann:1978rj}}
 It has been suggested by a number of authors
\cite{Uhlmann:1981aq,Uhlmann:1982fh,Jaffe:1989zv, Gibbons:1991ey,Gibbons:1993iv,Jaffe:2006uz,Jaffe:2007uy,Jaffe:2007kd,Anderson}          
that this construction may be generalised to a restricted class of
Riemannian manifolds,
including globally static spacetimes, which more generally 
 may be  characterised as ``Real Tunnelling Geometries''
\cite{Gibbons:1990ns}. 

One requires that the Riemann manifold admit a reflection symmetry
$\theta$, an involutive isometry which leaves fixed a two-sided hypersurface  
$\Sigma$ such  that
\ben
M= M_-\sqcup \Sigma \sqcup M_+ \,,\qquad \theta M_{\pm}= M_{\mp} \,, 
\een
where $\sqcup$ denotes the disjoint union.
Let $V$ be the space of complex valued functions that vanish outside $M_+$. 
If $|f \rangle \in V$   and ${\bar f}^\star_\theta = {\bar f}(\theta(x))$
is the pull-back of its complex conjugate under $\theta$,
we define 
\ben
\langle f| f\rangle = ||f\rangle|^2 = \int \int d^4 x \sqrt{g(x)} d^4 y \sqrt{g(y)}
  {\bar f}(\theta x) G(x,y)f(y) \,. \label{inner} 
\een
If the right hand side of (\ref{inner}) is positive definite
then the  left hand side provides a Hermitian inner product on
$V$ which may be identified with the  one-particle Hilbert space.

If
\ben
G_E= \frac{1}{(-\nabla^2 + m^2)} \,, 
\een
 then this reflection positivity condition 
is satisfied  \cite{Jaffe:2007kd}. However if 
\bea
G_E^{(4)}&=& \frac{1}{( -\nabla^2 + m_2^2    )(-\nabla ^2 + m_1^2)  }\\ 
&=& \frac{1}{m_1^2 - m_2^2 } \Bigl[  \frac{1}{(-\nabla^2 + m_2^2)} -
\frac{1}{(-\nabla^2 + m_1^2)}      \Bigr] \,. \label{partialfraction}
\eea
then reflection positivity is not satisfied\ \cite{Hawking:1985gh}.
To see this, suppose $m_1^2 > m_2^2 $ and  let
\ben
f=( -\nabla ^2 + m_2^2) h  
\een
where $h \in V$. Then
\bea
\langle f|f\rangle &=& \frac{1}{m_1^2 -  m_2^2}\int
\sqrt{g(x)} d^4x \int \sqrt{g(y)} d^4y
\Bigl [ {\bar f}(\theta x) h - 
  {\bar f}(\theta x) \frac{1}{(-\nabla^2+ m_1^2)}\, f(y)
  \Bigr ]\nonumber \\
&=& - \frac{1}{m_1^2 - m^2_2} \int \sqrt{g(x)} d^4x\int\sqrt{g(y)} d^4y
            {\bar f}(\theta x) \frac{1}{(-\nabla ^2 + m_1^2)}\, f(y)\,,
\eea
since  $ {\bar f}(\theta x)$ and $h$ have disjoint support.
Thus states of the form $|(-\nabla ^2 + m_2^2  ) h  \rangle $
have negative norm squared.

\subsection{$2N$'th-Order Case}

 For the $2N$'th-order operator we have
\be
\prod_{i=1}^N \fft1{(-\nabla^2 + m_i^2)} = \sum_{i=1}^N \fft{1}{C_i}\, 
\fft{1}{(-\nabla^2 + m_i^2)}\,,
\ee
where
\be
C_i=\prod_{j\ne i} (m_j^2-m_i^2)\,.
\ee
Proceeding as before, and considering
\be
f_i= \prod_{j\ne i} (-\nabla^2+m_j^2)\, h\,,
\ee
we shall have
\be
\langle f_i | f_i\rangle = \fft1{C_i}\, \int \sqrt{g(x)} d^4x\, 
 \int \sqrt{g(y)} d^4y\, \bar f_i(\theta x)\, \fft1{(-\nabla^2 + m_i^2)}\,
  f_i(y)\,.
\ee
This can have either sign, depending on the sign of $C_i$, and this
depends upon the relative values for the masses
$m_i$.  For example, if we order the masses so that
\be
m_1^2 > m_2^2 > m_3^2> \dots > m_N^2\,,
\ee
then $C_N$ is positive, and the signs of the $C_i$ alternate as $i$ 
decreases.  Thus there will always be states of negative norm as well as
states of positive norm.

  It is interesting to note that if we construct the static
Green function $G^{(2N)}(\br)$ generalising 
(\ref{greenf}) to the $2N$'th-order
operator, we obtain
\be 
G^{(2N)}(\br) =
\fft1{4\pi r}\, \sum_{i=1}^N \fft{e^{-m_i r}}{C_i}\,,\label{greenfN}
\ee
and that this is finite at $r=0$ for all $N\ge 2$.  In fact, one can see
that\footnote{These results can be proved by considering the contour
integrals
\bea
\oint_C dz\, z^p\, \prod_{i=1}^N \fft1{(z-m_i^2)}\nn
\eea
for $0\le p\le N-1$, where the contour $C$ is taken to be a circle of
radius $R$ centred on the origin, in the limit where $R$ is sent to infinity.}
\be
\sum_{i=1}^N \fft{m_i^{2p}}{C_i}=0\,,\ \ \hbox{for}\ \ p\le N-2\,,
\label{summC0}
\ee
and that
\be
\sum_{i=1}^N \fft{m_i^{2N-2}}{C_i}= (-1)^{N-1}\,.\label{summC1}
\ee
This implies that the Taylor expansion of $G^{(2N)}(\br)$ 
in (\ref{greenfN}) has the
form
\bea
G^{(2N)}(\br) &=& a_0 + a_2 \, r^2 + a_4\, r^4 + a_6\, r^6+\cdots\nn\\
 && + \fft{(-1)^{N-1}}{4\pi\, (2N-2)!}
\, r^{2N-3} + a_{2N-1}\, r^{2N-1} +a_{2N+1}\, r^{2N+1} +
\cdots\,,\label{taylorN}
\eea
where the constant $a_k$ coefficients depend on the masses $m_i$.  The
non-analyticity of the Green function, as a function of the Cartesian
coordinates, is associated with the occurrence of odd-integer powers
of $r$ in the Taylor expansion (\ref{taylorN}).  The first such term,
for the $2N$'th-order operator, occurs at the order $r^{2N-3}$.

  This softening of the Green function at short distance is  
analogous to the softening encountered in Pauli-Villars 
regularisation \cite{pauvil}. 

  The Euclidean Green function $G_E^{(2N)}(\br)$ 
for the $2N$'th-order Pais-Uhlenbeck 
operator can also be calculated using the same formalism as above.  From
(\ref{Geuc}), we see that
\be
G_E^{(2N)}(\br) = \sum_{i=1}^N \fft{m_i\, K_1(m_i\, r)}{4 \pi^2\, C_i\, r}\,.
\ee
The identities (\ref{summC0}) and (\ref{summC1}) again imply that
the non-analytic behaviour of $G_E^{(2N)}(\br)$ (in the form of the
$\log(m_i\, r/2)$ terms) is deferred to an increasingly higher
order in a small-$r$ expansion as $N$ increases.

\subsection{Higher Dimensional Green Functions}

  The Green function $G_E(\br)$ in $d$-dimensional Euclidean space, obeying
$(-\nabla^2+m^2) G_E(\br)=\delta(\br)$, is 
given by
\be
G_E(\br) = \fft{ m^{\fft{d}{2}-1}\, K_{\fft{d}{2}-1}(m r)}{
               (2\pi)^{\fft{d}{2}}\, r^{\fft{d}{2}-1} }\,. 
\ee
The cases we discussed previously correspond to $d=3$, for the static
Green function in four-dimensional spacetime, and $d=4$, for the
four-dimensional Euclidean Green function.  The analysis of the
Euclidean Green functions for the higher-order Pais-Uhlenbeck operators
proceeds in the same manner as in the cases we discussed previously.
As in those cases, the onset of non-analytic 
behaviour in the Green functions for the higher-order 
Pais-Uhlenbeck operators, namely
\be
G^{(2N)}_E(\br) = \sum_{i=1}^N
  \fft{ m_i^{\fft{d}{2}-1}\, K_{\fft{d}{2}-1}(m_i\, r)}{
               (2\pi)^{\fft{d}{2}}\, C_i\,r^{\fft{d}{2}-1} }\,,
\ee
 is deferred to increasingly 
higher powers in the small-$r$
expansion as $N$ increases.

\section{No-Hair Theorems for Static and Stationary Solutions}

  In this section, we address the question of whether there could exist
regular static solutions to certain classes of higher-derivative
field theories.  First, we consider field theories of the
Pais-Uhlenbeck type.  Then, we consider a class of $2N$ order field
theories with a general potential $V(\phi)$.

\subsection{Pais-Uhlenbeck Field Theories}\label{punohairsec}

Our discussion here applies rather generally to the situation where 
we have a static asymptotically flat background metric,
regular outside  a regular event horizon, in which
a scalar field which is also static satisfies the Pais-Uhlenbeck
equations 
\be
\prod_{i=1}^N (\Box-m_i^2)\, \phi=0\,.\label{NorderPU}
\ee
For concreteness, we shall first consider the fourth-order example
(\ref{EOM}).
If we integrate over the  domain $D$ obtained  by taking
an initial  spacelike surface $\Sigma_{(i)}$ extending from infinity to 
the horizon, and moving it up the orbits
of the static Killing vector field  $K^\mu$ which coincides on the horizon 
with its  null generators,
to a final surface  $\Sigma_{(f)}$,
and then use (\ref{ids}) with $\chi=\phi$, we find that
\bea
&&
 \half \int \Bigl(   (\nabla ^2 \phi)^2 + (m_1^2 + m_2^2 )
\, g ^ {\mu \nu}  \p _\mu \phi \p _\nu   \phi
+ m_1^2 m_2^2 \phi ^2       \Bigr )\sqrt{-g}  d \,^4 x \nn
\\ & =&  \int _{\p D}  \Bigl( - \phi (\nabla ^\mu \nabla ^2 \phi
+ (\nabla ^\mu   \phi) ( \nabla ^2 \phi) +(m_1^2 +m_2^2 )
 \phi \nabla ^\mu \phi \Bigr ) d \Sigma_\mu  \,. \label{ID}
\eea

Now $\p D$ has four boundary components: One is at infinity,
$\p D_\infty$; one is on the horizon, $\p D_H$; one is the initial surface
$\p D_i = \Sigma_{(i)}$; and the fourth is the  final
surface $ \p D_f= \Sigma_{(f)}$.
The integrals over $\p D_i $ and $\p D_f$ cancel one another, since, by
construction, $\Sigma_{(f)}$ is obtained by carrying $\Sigma_{(i)}$ along
the orbits of the Killing vector $K^\mu$.
The integrand on  $\p D_H$  vanishes because $d\Sigma_\mu$ on $H$ is
 proportional to $K^\mu$ and $K^\mu \p_\mu \phi= K^\mu \p _\mu
 \nabla ^2 \phi =0$.
The integral over $\p D_\infty$ vanishes by the boundary conditions on $\phi$.

Thus the left-hand side of (\ref{ID}) vanishes. But, unless $m_1^2m_2^2 =0 $, 
the integrand is positive definite. Thus in this case
$\phi=0$. If it happens that $m_1^2m_2^2 =0$ but $m_1^2 +m^2_2 \ne 0$,
we may only deduce that $\phi={\rm constant}$. If  $m_1^2 +m^2_2 =0$
we deduce that $\nabla ^2 \phi=0$ and hence, by a standard  argument,
that  $\phi={\rm constant}$ in this case as well.  Thus we have shown that
there can exist no regular, static solutions for the scalar field $\phi$ in
a static black hole background. A special case of this result is when there
is no black hole, and the spacetime is just flat Minkowski spacetime.

   The above argument generalises immediately to the $2N$-order
Pais-Uhlenbeck theory (\ref{NorderPU}), since by appropriate integrations by
parts, the analogue of the left-hand side in (\ref{ID}) will consist of
a sum of squares with positive coefficients.  The analogue of the surface terms on
the right-hand side will again vanish, by virtue of the assumption of
staticity and the imposition of an appropriate boundary condition on $\phi$ at
infinity.  Thus again, one concludes that there can be no non-trivial static
solutions for $\phi$ in flat space or a regular static black hole background.

  The result also immediately generalises to higher spacetime dimensions.

  One may also consider adding a scalar potential $V(\phi)$.  If this is
such that $\phi\, V'(\phi)$ is everywhere non-negative, then one
will again be able to show that $\phi$ must vanish for a static solution.

\subsection{Higher-Order Theories with a General Potential}\label{nhstatic2}

  In this section, we suppose that $\phi$ is static and obeys
\be
- (-\Box)^n\, \phi= V'(\phi)
\ee
in a curved $d$-dimensional spacetime. Differentiating with respect to
$\del_\mu$ and then contracting with $\del^\mu\phi$ implies
\be
-\del^\mu\phi\, \del_\mu (-\Box)^n\, \phi = V''\, g^{\mu\nu}\, 
\del_\mu\phi\, \del_\nu\phi\,.
\ee
By extracting derivatives sequentially, this can be written in the form
\be
\nabla_\mu S^\mu_{(n)} -(\Box^{\ft12(n+1)}\, \phi)^2
   = V''\, g^{\mu\nu}\, 
\del_\mu\phi\, \del_\nu\phi  \label{bochodd}
\ee
if $n$ is odd, or as
\be
\nabla_\mu S^\mu_{(n)} - (\nabla_\mu\, \Box^{\ft12 n}\,\phi)^2 =
V''\, g^{\mu\nu}\, \del_\mu\phi\, \del_\nu\phi\label{bocheven}
\ee
if $n$ is even.  

   If we assume that the solution $\phi$ is static, and that
the metric is static, then we can proceed as in section \ref{punohairsec},
by integrating (\ref{bochodd}) or (\ref{bocheven}) over a portion of
spacetime bounded by initial and final surfaces $\Sigma_{(i)}$
and $\Sigma_{(f)}$; an horizon $H$ and a boundary at infinity.  For the same
reasons as in section \ref{punohairsec}, the boundary contributions on
$\Sigma_{(i)}$ and $\Sigma_{(f)}$ will cancel, the term on the horizon will
vanish, and the contribution at infinity will give zero subject to 
appropriate fall-off conditions for $\phi$.  We may then conclude that if
potential is convex, i.e. $V''\ge 0$ everywhere, then $\phi$ must be a constant.

   By a straightforward extension of the argument above, we can similarly
conclude that there are no non-trivial static solutions to the equation
\be (-1)^{N+1}\, F(\Box)\, \phi = V'(\phi)\,,
\ee
where
\be
F(\Box) = \prod_{i=1}^N (\Box - m_i^2)\,.
\ee

\subsection{Virial Theorem}

   A different technique for ruling out static solutions under certain
assumptions about the potential is to consider the virial theorem.

   Calculating the spatial trace $T_{ii}$ of $T_{\mu\nu}$,
given by (\ref{Tmunu}), in a $d$-dimensional 
Minkowski background for a purely static field, we find that
\be
T_{ii}= \ft12 (2n+1-d)\, |Y|^2 - (d-1)\, V + \, \hbox{t.d.}\,,\label{Tii}
\ee
where t.d. denotes total derivatives, and 
\be
|Y|^2 = (\Box^{n/2}\phi)^2\,,\quad n=\, \hbox{even}\,;\qquad
|Y|^2 = (\del_i\Box^{(n-1)/2}\phi)^2\,,\quad n=\, \hbox{odd}\,.
\ee
(Because of the assumption that $\phi$ is static, $\Box$
is just $\del_i\del_i$ here.)
Similarly, the energy density is given
by
\be
T_{00}= \ft12 |Y|^2 +V\ + \,\hbox{t.d.}
\ee
Note that the total energy $\int d^{d-1}x\,  T_{00}$ is always 
positive provided that $V$ is non-negative.

   Because, for a general energy-momentum tensor, 
$\del_i T_{ij}=0$ for a static field in Minkowski spacetime, we have
$\del_i(x_j\, T_{ij})=T_{ii}$.  Integrating this over the $(d-1)$-dimensional
space, and assuming that the fall-off conditions for $\phi$ are
such that the surface term gives zero, it follows that we must have
\be
\int d^{d-1} x\, T_{ii} =0\,,
\ee
and hence, from (\ref{Tii}),
\be
\int d^{d-1}x\, \Big[ \fft12 (d-1-2n)\, |Y|^2 +(d-1) V\Big]=0\,.
\label{YV}
\ee
(The same result may be obtained by using Derrick's scaling argument
\cite{Derrick}, and
demanding that the total energy $\int d^{d-1}x\, T_{00}$ be extremised
for any static solution.)

  We see from (\ref{YV}) that if $V$ is positive and $d>2n+1$, there can
be no static solutions.  Similarly, if $V$ is negative
and $d<2n+1$, there can again be no static solutions.

\subsection{Stationary Metrics}

   The results of sections \ref{punohairsec} and \ref{nhstatic2} remain valid for
globally stationary metrics; that is, metrics admitting an everywhere timelike
Killing vector field $K^\mu$.  This is because the condition 
$K^\mu\, \del_\mu\phi=0$ implies that $\del_\mu\phi$ is spacelike, or zero,
and so $g^{\mu\nu}\, \del_\mu\phi\, \del_\nu\phi\ge0$.  In the case that
$K^\mu$ is not everywhere timelike, in other words an ergoregion is present, 
the argument just given will not necessarily be valid.  However, if the metric
is both stationary and axisymmetric, and $\phi$ is assumed to be both 
stationary and axisymmetric, the argument will go through.

\section{Stabilisation By Hubble Friction}

   Following the work of \cite{cosmo}, we have studied the behaviour 
of the Pais-Uhlenbeck oscillator in a background de Sitter universe.  We
choose coordinates in which the metric is given by
\be
ds^2= -dt^2 + e^{2Ht}\, dx_i dx_i\,,
\ee
where $H=\sqrt{\Lambda/3}$ and $\Lambda$ is the cosmological constant. The
non-linear 4th-order Pais-Uhlenbeck equation of motion for a spatially
homogeneous scalar field $\phi(t)$ with a potential $V(\phi)$ is
\be
\Big(\fft{d^2}{dt^2} + 3H \fft{d}{dt} + m_1^2\Big)
\Big(\fft{d^2}{dt^2} + 3H \fft{d}{dt} + m_2^2\Big)\phi = -V'(\phi)\,.
\label{FRWPU}
\ee
In the absence of the potential, we have two uncoupled damped simple
harmonic oscillators, with the solution
\be
\phi= \sum_{i=1}^2 A_i\, e^{-\fft32 Ht}\, \sin\Big(\sqrt{m_i^2-\fft{9H^2}{4}}\,
t + \alpha_i\Big)\,,
\ee
as long as $m_i^2 > 9H^2/4$. 

If we first set $H=0$ and consider, as an example, 
a potential $V=\fft14 \lambda\, \phi^4$,
we can verify Pavsic's observation \cite{pavsic1} 
that for given initial data, 
the solution in Minkowski spacetime 
remains bounded provided that $\lambda$ is sufficiently small.
If we choose a larger value of $\lambda$ such that the solution becomes
unstable, we can then verify, by numerical analysis, 
that turning on $H$ can render it stable again,
and in fact $\phi$ then decays to zero at large $t$ (i.e. within a few Hubble
times). 

One may also consider inhomogeneous solutions of the free 4th-order 
equation, with wave vector ${\bf k}$.  The equation of motion is obtained
by replacing $m_i^2$ by $m_i^2 + {\bf k}^2\, e^{-2Ht}$ in (\ref{FRWPU}).
The additional term decays to zero rapidly, and so we would expect that
our results in the non-linear situation will remain valid.  A complete
treatment would require further investigation going beyond the ODEs we
studied here.
 
Our results will be qualitatively similar for all the higher-order
Pais-Uhlenbeck oscillators.

\section{Conclusion}

In this paper we have studied a scalar field $\phi$ coupled to a 
fixed background metric $g_{\mu\nu}$, for which the scalar equation of motion
contains derivatives of arbitrary order.
In all the examples considered, the equations of
motion are linear in $\phi$  apart from
non-linearities $V'(\phi)$ introduced by a possible potential 
term $V(\phi)$.
It follows that classically the propagation is causal, that is,  
the characteristic surfaces across which discontinuities
may propagate are null hypersurfaces with respect to the 
metric $g_{\mu \nu}$. This is despite the fact that the various
energy-momentum tensors, such as the canonical or the 
Belinfante-Rosenfeld energy-momentum tensor, do not satisfy the
dominant energy condition.

We have obtained
general expressions for both the canonical and the
symmetric Belinfante-Rosenfeld energy-momentum tensors, in the latter
case by introducing auxiliary fields.  This formulation greatly 
simplifies the calculations, since it eliminates the
need to vary the metrics in high-order covariant derivatives of the
scalar field.
The Belinfante-Rosenfeld energy-momentum tensor is required 
for a consistent coupling of the scalar field
to the gravitational
field. We have also obtained general expressions for the conserved symplectic
current used in quantising the scalar field.
Both exhibit the well known difficulty that the energy-momentum tensors
will not not satisfy the usual energy conditions, and the 
sesquilinear inner product defined using the symplectic current
will not be positive definite.

Thus, if the standard definition of ``positive frequency" is adopted,
the quantum mechanical norm on states will be indefinite. On the other hand,
swapping  the sign in the definition of ``positive frequency"
for the modes carrying negative energy results in physical states carrying 
negative energy.

To investigate this further we have
considered Euclidean formulations of the quantised theory in
which the metric $g_{\mu \nu}$ is taken to be positive
definite. Assuming that the Riemannian manifold
admits a reflection map decomposing the manifold into
two disjoint domains symmetric about a separating hypersurface, we 
have investigated whether Osterwalder and Schrader's
 Reflection Positivity  condition holds. We show that this is never 
the case for the theories we study, confirming
earlier studies indicating that the Euclidean approach to these theories 
leads inevitably to  the presence of ``ghosts," that is states 
of negative norm.  

Finally, we studied, without assuming spherical symmetry,
the possible existence of non-singular finite energy static solutions of 
our equations on static or stationary backgrounds, and were able to rule out
such solutions subject to various assumptions on the  potential
$V(\phi)$. These results encompass the important cases of soliton 
solutions in flat space, bounce solutions
inducing the decay of false vacuua, and scalar hair for black holes. 

Although the implications of higher-derivative theories for quantum gravity
at the microscopic level remain unclear, 
the work of \cite{smilga1,smilga2,Smilga:2008pr,pavsic1,pavsic2}, 
and the cosmological
studies of \cite{cosmo}, indicate that there is still much to be understood
about the behaviour of such higher-derivative theories at the macroscopic,
classical, level.  After all, the discoveries of Slipher, Friedmann, 
Lema\^ itre and Hubble, Riess et al. \cite{frie,slip,lema,hubb,riess} clearly
indicate that the universe at the largest scale is definitely not in a 
state of static equilibrium.  Moreover, as we illustrated in this paper,
the phenomenon of Hubble friction
clearly mitigates against the instability, in the expanding state, of modes
that are unstable in a static configuration.

\section*{Acknowledgements}

We thank Stam Nicolis, Michael Volkov and Claude Warnick
for informative discussions on higher-derivative theories.
G.W.G. is grateful for the award of a LE STUDIUM Chair held at the  LMPT
of the University of Tours
under the auspices of which some of the work described in this paper
was carried out.  The work of C.N.P. is supported in part by DOE 
grant DE-FG02-13ER42020.  The research of S.S. was partially  
supported by the ERC-AdG-2015 grant 694896.  
C.N.P. is grateful the University of Tours for
hospitality during the course of some of this work.  S.S.  would like to 
thank M. Shaposhnikov for warm hospitality at EPFL during some stages of 
this project.

\appendix

\section{Canonical Noether Energy-Momentum Tensor}

   In this appendix, we discuss the construction of the canonical
conserved energy-momentum tensor for a scalar field with a higher-derivative
equation of motion, in a Minkowski 
spacetime background.  We show also in a some example cases how the
canonical energy-momentum tensor is related to the conventional 
energy-momentum that derived by the Belinfante-Rosenfeld procedure of
coupling the scalar to gravity and then varying the action with respect
to the metric.  In particular, we show in these examples how the 
canonical energy-momentum tensor can be derived by applying the 
Belinfante-Rosenfeld procedure to an action in which an appropriate
non-minimal coupling of the scalar field to gravity has been added.

   Consider a scalar field $\phi$ in Minkowski spacetime, described by a 
Lagrangian 
\be
{\cal L}= {\cal L}(\phi,\phi_{\nu_1},\phi_{\nu_1\nu_2},
\phi_{\nu_1\nu_2\nu_3},\cdots)\,,
\ee
where 
$\phi_{\nu_1}=\del_{\nu_1}\phi$, 
$\ \ \phi_{\nu_1\nu_2}=\del_{\nu_1}\del_{\nu_2}\phi$, etc.  The Euler-Lagrange
equation is
\be
\fft{\del\cL}{\del\phi}= \del_{\nu_1}\, \fft{\del\cL}{\del\phi_{\nu_1}} -
 \del_{\nu_1} \del_{\nu_2}\, \fft{\del\cL}{\del\phi_{\nu_1\nu_2}}
+ \del_{\nu_1} \del_{\nu_2}\del_{\nu_3}\, 
    \fft{\del\cL}{\del\phi_{\nu_1\nu_2\nu_3}} +\cdots\,.
\ee
Using this to substitute for $\del\cL/\del\phi$ in 
\be
\del_\mu\cL = \fft{\del\cL}{\del\phi}\, \phi_\mu + 
  \fft{\del\cL}{\del\phi_{\nu_1}}\, \phi_{\nu_1\mu} +
 \fft{\del\cL}{\del\phi_{\nu_1\nu_2}}\, \phi_{\nu_1\nu_2\mu} +
 \fft{\del\cL}{\del\phi_{\nu_1\nu_2\nu_3}}\, \phi_{\nu_1\nu_2\nu_3\mu}
+\cdots\,,
\ee
then after some straightforward manipulations one finds that
\be
\del_\mu\cL= \del_\nu\, \sum_{n\ge 1} W_\mu^{\sst{(n)}\, \nu}
\ee
and hence that $T_\mu{}^\nu$ defined by
\be
 T_\mu{}^\nu= -\sum_{n\ge 1} W_\mu^{\sst{(n)}\, \nu} + \delta_\mu^\nu\, \cL
\label{canonT}
\ee
is conserved, $\del_\nu\, T_\mu{}^\nu=0$, where
\bea
W_\mu^{\1\, \nu}&=& \fft{\del\cL}{\del\phi_\nu}\, \phi_\mu\,,\nn\\
W_\mu^{\2\, \nu}&=& \fft{\del\cL}{\del\phi_{\alpha_1\nu}}\, 
  \phi_{\alpha_1\mu} -\del_{\alpha_1}\, \fft{\del\cL}{\del\phi_{\alpha_1\nu}}\,
 \phi_\mu \,,\nn\\
W_\mu^{\3\, \nu}&=& \fft{\del\cL}{\del\phi_{\alpha_1\alpha_2\nu}}\, 
  \phi_{\alpha_1\alpha_2\mu} -\del_{\alpha_1}\, 
 \fft{\del\cL}{\del\phi_{\alpha_1\alpha_2\nu}}\, \phi_{\alpha_2\mu}  
  +\del_{\alpha_1}\del_{\alpha_2}\, 
 \fft{\del\cL}{\del\phi_{\alpha_1\alpha_2\nu}}\, \phi_\mu\,,\nn\\
W_\mu^{\4\, \nu}&=& \fft{\del\cL}{\del\phi_{\alpha_1\alpha_2\alpha_3\nu}}\,
  \phi_{\alpha_1\alpha_2\alpha_3\mu} -\del_{\alpha_1}\, 
 \fft{\del\cL}{\del\phi_{\alpha_1\alpha_2\alpha_3\nu}}\, 
  \phi_{\alpha_2\alpha_3\mu}         
  +\del_{\alpha_1}\del_{\alpha_2}\,
 \fft{\del\cL}{\del\phi_{\alpha_1\alpha_2\alpha_3\nu}}\, \phi_{\alpha_3\mu}
\nn\\
&& -\del_{\alpha_1}\del_{\alpha_2}\del_{\alpha_3}\, 
  \fft{\del\cL}{\del\phi_{\alpha_1\alpha_2\alpha_3\nu}}\, \phi_\mu\,,
\eea
and so on.  Note that $-W_\mu^{\sst{(n)}\, \nu}$ gives the contribution
to the canonical energy-momentum tensor associated with the 
dependence of the Lagrangian on the $n$th-order spacetime derivatives of $\phi$.
As always with the canonically-defined energy-momentum tensor coming from 
the Noether symmetries of the Lagrangian (i.e. the fact that the
Lagrangian has no explicit dependence on the spacetime coordinates $x^\mu$),
$T_\mu{}^\nu$ may not necessarily be symmetric in $\mu$ and $\nu$ after
lowering the $\nu$ index.  In the event that it is not, one can construct
a conserved symmetric 2-index tensor from it by making use of the freedom
to add a term $\del_\sigma\, \psi_\mu{}^{\nu\sigma}$ to $T_\mu{}^\nu$,
where $\psi_\mu{}^{\nu\sigma}$ is any tensor that is antisymmetric
in $\nu$ and $\sigma$,
\be
\psi_\mu{}^{\nu\sigma} = -\psi_\mu{}^{\sigma\nu}\,.
\ee

   Here we present examples of the canonical energy-momentum tensors
for some simple Lagrangians.  For the standard kinetic term
for a scalar field we obtain
\bea
\cL &=& -\fft12 (\del\phi)^2\,,\nn\\
T_\mu{}^\nu &=& \del_\mu\phi\, \del^\nu\phi -\fft12 \delta_\mu^\nu\, 
   (\del\phi)^2\,,
\eea
which is the same as the conventional Belinfante-Rosenfeld 
energy-momentum tensor that one would
obtain from varying the metric in $\cL=-\ft12 \sqrt{-g}\, (\del\phi)^2$ and
then specialising to a Minkowski background.
If instead we were to integrate the Lagrangian by parts before the
calculation of the canonical energy-momentum tensor, we would obtain
instead the energy-momentum tensor $\widetilde T_\mu{}_\nu$:
\bea
\cL &=& \fft12 \phi\, \Box\, \phi\,,\nn\\
\widetilde T_\mu{}^\nu &=& -\fft12\phi\, \del_\mu\del^\nu\phi +
  \fft12 \del_\mu\phi\, \del^\nu\phi  +\fft12 \delta_\mu^\nu\,
   \phi\,\Box\,\phi\,.\label{T2c}
\eea
The difference between the two is
\be
\widetilde T_\mu{}^\nu-  T_\mu{}^\nu =
  \fft14 \Big(\delta_\mu^\nu \,\Box - \del_\mu \del^\nu\Big)\, \phi^2\,,
\ee
and in fact $\widetilde T_\mu{}^\nu$ is the conventional 
Belinfante-Rosenfeld energy-momentum tensor one would obtain by varying the
metric in the non-minimally coupled Lagrangian
\be
\widetilde\cL = 
\sqrt{-g}\, \Big[ -\fft12 (\del\phi)^2 - \fft18 R\, \phi^2\Big]\,,
\ee
and then specialising to the Minkowski spacetime background.

  If we consider now a 4th-order Lagrangian, we obtain
\bea
\cL &=& -\fft12 (\Box\phi)^2\,,\nn\\
T'_\mu{}^\nu &=& (\Box\phi)\, \del_\mu \del^\nu\phi - (\del^\nu\Box\phi)\,
\del_\mu\phi - \fft12 (\Box\phi)^2\, \delta_\mu^\nu\,.
\eea
We have denoted this tensor with a prime, because it is not symmetric 
in $\mu$ and (the lowered) $\nu$ indices.  We can
obtain a symmetric conserved energy-momentum tensor by adding a term
$\del_\sigma\, \psi_\mu{}^{\nu\sigma}$ with
\be
\psi_\mu{}^{\nu\sigma}= (\del_\mu\del^\nu\phi)\, \del^\sigma\phi -
                         (\del_\mu\del^\sigma\phi)\, \del^\nu\phi\,,
\ee
leading to the energy-momentum tensor $T_\mu{}^\nu=T'_\mu{}^\nu+ 
  \del_\sigma\, \psi_\mu{}^{\nu\sigma}$, where 
\bea
T_\mu{}^\nu &=& 2(\p_\mu \p^\nu \phi) \Box \phi -
(\p_\mu \phi  ) \Box (\p ^\nu \phi) - (\p^\nu \phi) \Box (\p_\mu \phi)
+(\p_\mu \p^\nu \p_\alpha \phi)( \p^\alpha \phi) \nn\\
&& -
( \p_\mu \p ^\alpha \phi ) (\p ^\nu \p_\alpha \phi )
-\half \delta_\mu^\nu\, (\Box \phi )^2\,.\label{T4c}
\eea
The tensors (\ref{T2c}) and (\ref{T4c}) are precisely of the
form of the 2nd-order and 4th-order contributions in the 
energy-momentum tensor (\ref{T1}) given in \cite{Andrzejewski:2009bn}
for the 4th-order Pais-Uhlenbeck theory.

  The canonical energy-momentum tensor $T_\mu{}^\nu$ for the 
$-\ft12(\Box\phi)^2$ Lagrangian, given by (\ref{T4c}), is different from the
flat-space specialisation of the 
expression (\ref{4thorderT}) that we obtained by the 
Belinfante-Rosenfeld metric-variation method. In fact the difference between 
the two energy-momentum tensors can be generated as the metric 
variation of a certain non-minimal coupling of the scalar field
to gravity.  Specifically, we find that the canonical energy-momentum
tensor in (\ref{T4c}) can be obtained by varying the metric in the Lagrangian
\be
\widetilde\cL = \sqrt{-g}\, \Big[ -\fft12 (\Box\phi)^2 +
  (R_{\mu\nu} - \fft14 R \,g_{\mu\nu})\, \del^\mu\phi\, \del^\nu\phi\Big]\,,
\label{4thnm}
\ee
and then specialising to a Minkowski-spacetime background.  The non-minimal
coupling term in (\ref{4thnm}) is similar in form to a Horndeski
coupling \cite{Horndeski:1974wa} that has been studied in other contexts 
in the literature,
namely a term $\sqrt{-g}(R_{\mu\nu} -\ft12 R g_{\mu\nu})\, \del^\mu\phi\,
\del^\nu\phi$.

\end{document}